\documentclass{article}
\usepackage[utf8]{inputenc}
\usepackage[margin=1in]{geometry}
\usepackage{graphicx}
\usepackage{natbib}
\usepackage{amsmath, amssymb, bm, amsthm}
\usepackage{boxedminipage}
\usepackage{multirow}
\usepackage{url}

\newtheorem{assump}{Assumption}

\renewcommand{\vec}{\mathrm{vec}}
\newcommand{\tr}{\mathrm{tr}}

\DeclareMathOperator*{\argmin}{arg\,min}

\title{\bf Hybrid Kronecker Product Decomposition and Approximation}
\author{Chencheng Cai, Rong Chen and Han Xiao\footnote{Chencheng Cai is a Ph.D student
    at Department of Statistics, Rutgers
    University, Piscataway, NJ 08854. E-mail:
    chencheng.cai@rutgers.edu.
Rong Chen is Professor, Department of
    Statistics, Rutgers University, Piscataway, NJ 08854. E-mail:
    rongchen@stat.rutgers.edu. Han Xiao is Associate Professor,
    Department of Statistics, Rutgers
    University, Piscataway, NJ 08854. E-mail:
    hxiao@stat.rutgers.edu.
Rong Chen is the
    corresponding author. Chen's research is supported
in part by National Science Foundation
grants DMS-1503409, DMS-1737857, IIS-1741390 and
CCF-1934924.}
\\ \vspace{0.2cm} {Rutgers University}\\}
\date{}

\begin{document}

\maketitle

\begin{abstract}
  Discovering the underlying low dimensional structure of high dimensional
  data has attracted a significant amount of researches recently and
  has shown to have a wide range of applications. 
 As an effective dimension reduction tool, singular value decomposition is often used to analyze high dimensional matrices, which are traditionally assumed to have a low rank matrix approximation.
  In this paper, we propose a new approach. 
  We assume a high dimensional matrix can be approximated by a sum of a small
  number of Kronecker products of matrices
  with potentially different configurations,
  named as a {\it h}ybird {\bf K}ronecker {\bf o}uter {\bf P}roduct
  {\bf A}pproximation ($h$KoPA).
  It provides an extremely flexible way of
  dimension reduction compared to the low-rank matrix approximation.  
  Challenges arise in estimating a $h$KoPA
  when the configurations of component Kronecker products are
  different or unknown. We propose an estimation procedure when the set of
  configurations are given and a joint configuration determination and component estimation procedure
  when the configurations are unknown. Specifically, a least squares
  backfitting algorithm is used when the configuration is given. When the
  configuration is unknown, an iterative greedy algorithm is used.
  Both simulation and real image examples show that the proposed algorithms
  have promising performances. The hybrid Kronecker product approximation 
  may have potentially wider applications in low dimensional
  representation of high dimensional data.
\end{abstract}

\newpage

\section{Introduction}
High dimensional data often has low dimensional structure that allows
significant dimension reduction and compression. In applications such as
data compression, image denoising and processing, matrix completion,
high dimensional matrices
are often assumed to be of low rank and can be represented as a sum of several
rank-one matrices (vector outer products) in
a singular value decomposition (SVD) form.
\cite{Eckart1936the} reveals the connection between singular value decomposition and low-rank matrix approximation. Recent studies include 
image low-rank approximation \citep{freund1999short}, 
principle component analysis \citep{wold1987principal, zou2006sparse}, 
factorization in high dimensional time series \citep{lam2012factor, yu2016temporal},
non-negative matrix factorization \citep{hoyer2004non, cai2009locality},
matrix factorization for community detection \citep{zhang2012overlapping, yang2013overlapping, le2016optimization}, 
matrix completion problems \citep{candes2009exact, candes2010matrix, yuan2016tensor}, 
low rank tensor approximation \citep{grasedyck2013literature},
machine learning applications \citep{guillamet2002non, pauca2004text, zhang2008topology, sainath2013low},
 among many others.


As an alternative to vector outer product, the Kronecker product is another
way to represent a high dimensional matrix with a much fewer number of elements.
The decomposition of a high dimensional matrix into the
sum of several Kronecker products of identical configuration is known as
Kronecker product decomposition \citep{van1993approximation}. Here
{\it configuration} refers to the dimensions of the component
matrices of the Kronecker product.
The form of Kronecker product appears in many fields including signal processing, image processing and quantum physics  \citep{Werner2008On, Duarte2012Kronecker,Kaye2007introduction}, where the data has an intrinsic Kronecker product structure.
\cite{cai2019kronecker} considers to model a high dimensional matrix with a sum
of several Kronecker products of the same but unknown configuration. For a given configuration, the approximation using a
sum of several Kronecker products can be turned into an approximation using
a sum of several
rank-one matrices problem after a rearrangement operation of the
matrix elements \citep{van1993approximation, cai2019kronecker}. The unknown configuration can be estimated by minimizing an information criterion as proposed
by \cite{cai2019kronecker}.

However, it is often the case that the {\bf K}ronecker {\bf o}uter
{\bf P}roduct (KoPA) using a
single configuration requires a large number of terms to make the approximation
accurate. By allowing the use of a sum of Kronecker products of different
configurations, an observed high dimensional matrix (image)
can be approximated more effectively using a much smaller number of parameters (elements).
We note that often the observed matrix
can have much more complex structure than a single Kronecker product can handle.
For example, representing an image with Kronecker products of the same configuration
is often not satisfactory since the configuration dimensions determine the
block structure of the recovered image, similar to
the pixel size of the image. A single configuration is often not
possible to provide as much details as needed, as seen from the top row of Figure 10
later. 
Similar to the extension from low rank matrix approximation to KoPA of
a single configuration, we propose to extend the Kronecker product
approach to allow for multiple configurations. It is
more flexible and may provide more accurate representation
with a smaller number of parameters.

In this paper, we generalize the KoPA model in \cite{cai2019kronecker} to a multi-term setting, where the observed high dimensional matrix is assumed to be generated from a sum of several Kronecker products of different
configurations -- we name the model {\it hybrid} KoPA ($h$KoPA).
As a special case, when all the Kronecker products are vector outer products,
KoPA corresponds to a low rank matrix approximation.

We consider two problems in this paper. We first propose a procedure to
estimate a $h$KoPA with a set of known configurations.
The procedure is based on an iterative backfitting algorithm
with the basic operation
being finding the least squares solution of a single Kronecker product of a
given configuration to a given matrix. This operation
can be obtained through a rearrangement operation and a SVD
estimation.
Next, we consider the problem of determining the configurations in a
$h$KoPA for a given observed matrix.
As exploiting the space of all possible configuration combinations
is computationally expensive, we propose
an iterative greedy algorithm similar to the boosting algorithm \citep{freund1999short}.
In each iteration, a single Kronecker product term is added
to the model by fitting the residual matrix from the previous iteration.
The configuration of the added Kronecker product is determined
using the procedure proposed in \cite{cai2019kronecker}.
This algorithm efficiently fits a $h$KoPA model with a
potentially sub-optimal solution as a compromise between computation
and accuracy.

The rest of the paper is organized as follows. The $h$KoPA
model is introduced and discussed in Section \ref{sec:mtkpm}, with
a set of identifiability assumptions.
In Section \ref{sec:method}, we provide the details of the
iterative backfitting estimation procedure for the
model with known configurations and the greed
algorithm to fit a $h$KoPA with unknown configurations.
Section \ref{sec:example} demonstrates the performance of the proposed
procedures with a simulation study  and a real image example.
Section \ref{sec:conclusion} concludes.

\textbf{Notations}: For a matrix $\bm M$, $\|\bm M\|_F:=\sqrt{\tr(\bm M\bm M^T)}$ stands for its Frobenius norm and $\|\bm M\|_S$ stands for its spectral norm, which corresponds to the largest singular value of $\bm M$.

\section{Hybrid Kronecker Product Model}\label{sec:mtkpm}

For simplicity, in this paper we assume the dimensions $(P,Q)$ of the observed matrix are of powers of 2. As a consequence, all the component matrices in $h$KoPA are also of powers of 2. The model and procedures apply to all other cases of $(P,Q)$ for which the dimensions of the component matrices in $h$KoPA are then the factors of $P$ and $Q$, with more complex notations.

In a $K$-term $h$KoPA model, we assume that the observed
$2^M\times 2^N$ matrix $\bm Y$
is generated from the sum of $K$ Kronecker products of configurations
$(m_k, n_k)$,
\begin{equation}
    \bm Y = \sum_{k=1}^K\lambda_k\bm A_k\otimes \bm B_k + \sigma \bm E,\label{eq:mtkpm}
\end{equation}
where the matrix $\bm A_k$ is of dimension $2^{m_k}\times 2^{n_k}$,
the matrix $\bm B_k$ is of dimensions $2^{M-m_k}\times 2^{N-n_k}$,
for $k=1,\dots, K$. The matrix $\bm E$ is the noise matrix
with i.i.d. standard Gaussian entities. The operation $\otimes$ denotes the Kronecker product such that for any two matrices $\bm A\in\mathbb R^{2^m\times 2^n}$ and $\bm B\in\mathbb R^{2^{M-m}\times 2^{N-n}}$, $\bm A\otimes \bm B$ is a
$2^M\times 2^N$ block matrix defined by
$$\bm A\otimes \bm B = \begin{bmatrix}
a_{1, 1}\bm B& a_{1, 2}\bm B& \cdots& a_{1, 2^n}\bm B\\
a_{2, 1}\bm B& a_{2, 2}\bm B& \cdots& a_{2, 2^n}\bm B\\
\vdots & \vdots & \ddots & \vdots \\
a_{2^m, 1}\bm B& a_{2^m, 2}\bm B & \cdots & a_{2^m, 2^n}\bm B
\end{bmatrix},$$
where $a_{i, j}$ is the $(i, j)$-th element of $\bm A$. To better present the block structure of the Kronecker product, for any $2^M\times 2^N$ matrix $\bm M$,
we denote $\bm M_{i, j}^{p, q}$ as the $(i, j)$-th $2^p\times 2^q$
block of $\bm M$ such that $[\bm A\otimes \bm B]_{i, j}^{M-m, N-n} = a_{i, j}\bm B$.
We denote the dimensions $(m,n,M-m,N-n)$ as the configuration of the
Kroneker product $\bm A\otimes \bm B$. When $(M,N)$ are fixed, we simplify the
configuration notation as $(m,n)$.

We denote the configuration of the $h$KoPA model
in (\ref{eq:mtkpm}) as the collection of configurations
${\cal C}  = \{(m_k, n_k)\}_{k=1}^K$. The Kronecker product components
in (\ref{eq:mtkpm}) are allowed to have different configurations $(m_k, n_k)$.
When the model configuration ${\cal C}$ is known, we need to estimate
the component matrices $\bm A_k$ and $\bm B_k$, for $k=1,\ldots, K$
in model (\ref{eq:mtkpm}). When the configuration ${\cal C}$ is unknown,
the estimation of model (\ref{eq:mtkpm}) requires the determination of the
configuration ${\cal C}$ as well,
resulting in a configuration determination problem in addition to the
estimation problem.

Some existing researches on Kronecker product structured data can be viewed as
the special cases of model (\ref{eq:mtkpm}).
When $K=1$ and the configuration is unknown, model (\ref{eq:mtkpm}) reduces
to the single-term KoPA model investigated in
\cite{cai2019kronecker}. When the configurations of the $K$ Kronecker products
are known and equal such that $(m_1,n_1)=(m_2,n_2)=\dots = (m_K, n_K)$,
an estimation of model (\ref{eq:mtkpm}) is provided by the Kronecker product
decomposition approximation \citep{van1993approximation}.

First we note that model (\ref{eq:mtkpm}) is not identifiable. Similar to
the single-term KoPA model in \cite{cai2019kronecker}, Assumption \ref{assump:norm-1} below is imposed such that $\bm A_k$ and $\bm B_k$ are normalized and the Kronecker products are in decreasing order of the coefficient $\lambda_k$
such that given the values of the $K$ Kronecker products, $\lambda_k$, $\bm A_k$ and $\bm B_k$ can be uniquely determined with an exception for the sign changes of $\bm A_k$ and $\bm B_k$.

\begin{assump}[Identifiability Condition 1]\label{assump:norm-1}
We assume
$$\|\bm A_k\|_F=\|\bm B_k\|_F=1, \text{ for }k=1,\dots, K,$$
and $\lambda_1\geqslant\lambda_2\geqslant \cdots \geqslant \lambda_K >0$.
\end{assump}

Notice that if a matrix $\bm A_1$ of dimension $2^{m_1}\times 2^{n_1}$ is {\it smaller} than a matrix $A_2$ of 
 dimension $2^{m_2}\times 2^{n_2}$ in that $m_1 \leqslant m_2,\ n_1\leqslant n_2$, then for
any $2^{m_2-m_1}\times 2^{n_2-n_1}$ matrix $\bm C$, we have
\begin{equation}
\lambda_1\bm A_1\otimes (\bm B_1 + \lambda_2\bm C\otimes \bm B_2) + \lambda_2(\bm A_2 - \lambda_1\bm A_1\otimes \bm C)\otimes \bm B_2= \lambda_1\bm A_1\otimes \bm B_1 + \lambda_2\bm A_2\otimes \bm B_2,\label{eq:equivalence}
\end{equation}
for any $\bm B_1$ of dimension $2^{M-m_1}\times 2^{N-n_1}$, and $\bm B_2$ of dimension $2^{M-m_2}\times 2^{N-n_2}$.
Therefore, additional identifiability condition is needed to be imposed to
$\bm A_k$ and $\bm A_l$ when $m_k\leqslant m_l$ and $n_k\leqslant n_l$ as shown
in the following Assumption \ref{assump:orthogonal}.
\begin{assump}[Identifiability Condition 2]\label{assump:orthogonal}
  For any $0\leq k, l \leq K$ such that $m_k\leqslant m_l$ and
  $n_k\leqslant n_l$, we assume
$$\tr\left[\bm A_l(\bm A_k\otimes \bm 1_{i, j}^{m_l-m_k, n_l-n_k})^T\right]=0,$$
  for all $1\leq i \leq 2^{m_l-m_k}$ and $1\leq j \leq 2^{n_l-n_k}$,
  where $\bm 1_{i, j}^{m_l-m_k, n_l-n_k}$ denotes the $2^{m_l-m_k}\times 2^{n_l-n_k}$ matrix whose $(i, j)$-th element is 1 with all other elements being 0.
Furthermore, if $m_k=m_l$ and $n_k=n_l$, we assume
$$\tr[\bm A_l\bm A_k^T] = \tr[\bm B_l\bm B_k^T]=0.$$
\end{assump}
Assumption \ref{assump:orthogonal} is an orthogonal assumption on $\bm A_k$ and $\bm A_l$ when both dimensions of $\bm A_k$ are less than or equal to the ones of $\bm A_l$. If $\bm A_k$ and $\bm A_l$ do not satisfy the condition in Assumption \ref{assump:orthogonal}, one can always perform an orthogonalization operation by finding a $2^{m_l-m_k}\times 2^{n_l-n_k}$ non-zero matrix $\bm C$, whose $(i, j)$-th element is given by
\begin{equation}
[\bm C]_{i, j} = \tr\left[\bm A_l(\bm A_k\otimes \bm 1_{i, j}^{m_l-m_k, n_l-n_k})^T\right],\label{eq:C}
\end{equation}
such that Assumption \ref{assump:orthogonal} is satisfied for $\bm A_k$ and $\bm A^*_l = (\bm A_l - \bm A_k\otimes \bm C)/\|\bm A_l - \bm A_k\otimes \bm C\|_F$. Note that $\bm C$ in (\ref{eq:C}) is the least squares solution of $\min \|\bm A_l - \bm A_k\otimes \bm C\|_F^2$. The procedure of orthogonalizing $\bm A_k$ and $\bm A_l$ can be generalized to multiple terms through the Gram-Schmidt process depicted in Figure \ref{fig:gram-schmidt-process}. Note that the identifiability Assumption \ref{assump:orthogonal} can be replaced with the same
condition on the $\bm B$s, but there is no need to impose the condition on both
$\bm A$'s and $\bm B$'s.

\begin{figure}[!htbp]
    \begin{center}
        \begin{boxedminipage}{6.5in}
        \caption{Gram-Schmidt Process for $h$KoPA Model}
        \label{fig:gram-schmidt-process}
        \it{
        \begin{itemize}
            \item Let $(i_1,\dots, i_K)$ be a shuffle of the set $\{1, \dots, K\}$ such that
            \begin{enumerate}
                \item $m_{i_j}\leqslant m_{i_k}$ for all $j\leqslant k$,
                \item $n_{i_j}\leqslant n_{i_k}$ for all $j\leqslant k$ such that $m_{i_j}=m_{i_k}$.
            \end{enumerate}
            \item Set $\bm A_{i_1}^* = \bm A_{i_1}$, $\bm B_{i_1}^* = \bm B_{i_1}$ and $\lambda_{i_1}^*=\lambda_{i_1}$.
            \item For $j=2,\dots, K$,
            \begin{itemize}
                \item Let $\Omega_{j}=\{k< j: m_k\leqslant m_j, n_k\leqslant n_j\}$.
                \item Fit the following lease square problem
                $$\min_{\bm C_k^j\in\mathbb R^{2^{m_{i_j}-m_{i_k}}\times 2^{n_{i_j}-n_{i_k}}}} \|\bm A_{i_j}-\sum_{k\in\Omega_j}\bm A_{i_k}^*\otimes \bm C_{k}^j\|_F^2,$$
                and denote the solution as $\hat{\bm C}_k^j$ for $k\in\Omega_j$.
                \item Set
                \begin{align*}
                    \bm A_{i_j}^* &= (\bm A_{i_j} - \sum_{k\in\Omega_j}\bm A_{i_k}^*\otimes \hat{\bm C_k}^j)/\|\bm A_{i_j} - \sum_{k\in\Omega_j}\bm A_{i_k}^*\otimes \hat{\bm C_k}^j\|_F,\\
                    \lambda_{i_j}^* &= \lambda_{i_j}\|\bm A_{i_j} - \sum_{k\in\Omega_j}\bm A_{i_k}^*\otimes \hat{\bm C_k}^j\|_F,\\
                    \bm B_{i_j}^*&=\bm B_{i_j}.
                \end{align*}
                \item For $k\in\Omega_j$, update
                \begin{align*}
                    \bm B_{i_k}^* &\leftarrow \left(\bm B_{i_k}^* + \dfrac{\lambda_{i_j}}{\lambda^*_{i_k}}\hat{\bm C}_k^j\otimes \bm B_{i_k}^*\right)/\left\|\bm B_{i_k}^* + \dfrac{\lambda_{i_j}}{\lambda^*_{i_k}}\hat{\bm C}_k^j\otimes \bm B_{i_k}^*\right\|_F,\\
                    \lambda_{i_k}^* &\leftarrow \lambda_{i_k}^*\left\|\bm B_{i_k}^* + \dfrac{\lambda_{i_j}}{\lambda^*_{i_k}}\hat{\bm C}_k^j\otimes \bm B_{i_k}^*\right\|_F.
                \end{align*}
            \end{itemize}
            \item $\lambda_i^*, \bm A_i^*, \bm B_i^*$, $i=1,\dots, K$, satisfy Assumptions \ref{assump:norm-1} and \ref{assump:orthogonal}.
        \end{itemize}
         }
        \end{boxedminipage}
    \end{center}
\end{figure}

\section{Methodology}\label{sec:method}
In this section, we propose the estimation procedures for fitting the
$h$KoPA model in (\ref{eq:mtkpm}). Specifically,
when the configurations are known, we adopt the backfitting approach
(or alternating least squares approach) to fit the model.
And when the configurations are unknown, we propose a greedy approach
by adding one Kronecker product component at a time.

\subsection{Hybrid Kronecker Product Model with Known Configurations}
When the configurations $(m_k, n_k)$, $k=1,\dots, K$, are known, we consider the following least squares problem.
\begin{equation}
    \min \|\bm Y - \sum_{k=1}^K\lambda_k\bm A_k\otimes \bm B_k\|_F^2.\label{eq:least-square-optimization}
\end{equation}
When $K=1$, such a problem can be solved by singular value decomposition of a
rearranged version of matrix $\bm Y$ as shown in \cite{cai2019kronecker}.
Specifically, the rearrangement operation $\mathcal R_{m,n}[\cdot]$ reshape the
$2^M\times 2^N $matrix $\bm Y$ to a new $2^{m+n}\times 2^{M+N-m-n}$ matrix such
that
$$
\mathcal R_{m,n}[\bm Y] = [\vec(\bm Y_{1, 1}^{M-m, N-n}),\dots, \vec(\bm Y_{1, 2^n}^{M-m, N-n}), \dots, \vec(\bm Y_{2^m, 1}^{M-m, N-n}),\dots, \vec(\bm Y_{2^m, 2^n}^{M-m, N-n})]^T,
$$
where $\bm Y_{i, j}^{M-m, N-n}$ stands for the $(i, j)$-th $2^{M-m}\times 2^{N-n}$ block of matrix $\bm Y$ and $\vec(\cdot)$ is the vectorization operation that flattens a matrix to a column vector. It is pointed out by \cite{van1993approximation} and \cite{cai2019kronecker} that the rearrangement operation can transform a Kronecker product to a vector outer product such that
$$\mathcal R_{m,n}[\bm A\otimes \bm B]= \vec(\bm A)\vec(\bm B)^T.$$
This can be seen from the fact that all the elements in the matrix
$\bm A\otimes \bm B$
are in the form of $a_{i, j}b_{k,\ell}$, which is exactly the same as those in
$\vec(\bm A)\vec(\bm B)^T$, where $a_{i,j}$ is the $(i,j)$-th element in $\bm A$
and $b_{k,\ell}$ is the $(k,\ell)$-th element in $\bm B$.

Therefore, the least squares optimization problem
\[
\min \|\bm Y - \lambda\bm A\otimes \bm B\|_F^2,
\]
is equivalent to a rank-one matrix approximation problem since
$$\|\bm Y-\lambda\bm A\otimes \bm B\|_F^2 =
\|\mathcal R_{m,n}[\bm Y] - \lambda\vec(\bm A)\vec(\bm B)^T\|_F^2,$$
whose solution is given by the leading component in the singular value decomposition
of $\mathcal R_{m,n}[\bm Y]$ as proved by \cite{Eckart1936the}.

When there are multiple terms $K>1$ in the model (\ref{eq:mtkpm}), we propose
to solve the optimization problem (\ref{eq:least-square-optimization})
through a backfitting algorithm (or alternating least squares algorithm)
by iteratively estimating $\lambda_k$, $\bm A_k$ and $\bm B_k$ by
\[
\min \|\bm (Y - \sum_{i\neq k} \hat \lambda_i\hat {\bm A}_i\otimes \hat {\bm B}_i) -
\lambda_k\bm A_k\otimes \bm B_k\|_F^2,
\]
using the rearrangement operator and SVD, with fixed
$\hat \lambda_i$, $\hat {\bm A}_i$ and $\hat {\bm B}_i$ ($i\neq k$) from previous iterations.

When all configurations $\{(m_k, n_k)\}_{k=1}^K$ are distinct,
the backfitting procedure for $h$KoPA is depicted in
Figure \ref{fig:als}, where $\vec^{-1}_{m, n}$ is the inverse of the vectorization
operation that convert a column vector back to a $2^m\times 2^n$ matrix.
When $r$ terms indexed by $k_1,\dots, k_r$ in the $h$KoPA model have the same configuration, these terms are updated simultaneously in the backfitting algorithm by keeping the first $r$ components from the SVD of the residual matrix
$\bm E_k=\bm Y - \sum_{i\neq k_1,\dots, k_r} \hat \lambda_i\hat {\bm A}_i\otimes \hat {\bm B}_i$.
We also Orthonormalize the components by the Gram-Schmidt process at the end of each backfitting round.

Since each iteration of the backfitting procedure reduces
the sum of squares of residuals, the algorithm always converges, though it may
land in a local optimal. Empirical experiences show that most of the time
the global minimum is reached. Starting with different initial values and
order of backfitting helps.

\begin{figure}[!htbp]
    \begin{center}
        \begin{boxedminipage}{6.5in}
        \caption{Backfitting Least Squares Procedure}
        \label{fig:als}
        \it{
            \begin{itemize}
                \item Set $\hat\lambda_1=\hat \lambda_2=\dots=\hat\lambda_K=0$.
                \item Repeat until convergence:
                \begin{itemize}
                    \item For $k=1,\dots, K$,
                    \begin{itemize}
                        \item Let $\bm E_k = \bm Y - \sum_{i\neq k}\hat\lambda_i\hat{\bm A}_i\otimes \hat{\bm B}_i$.
                        \item Compute SVD of $\mathcal R_{m_k,n_k}[\bm E_k]$ such that
                        $$\mathcal R_{m_k,n_k}[\bm E_k] = \sum_{j=1}^Js_j\bm u_j\bm v_j^T,$$
                        where the singular values are in decreasing order such that $s_1\geqslant s_2\geqslant\dots\geqslant s_J $.
                        \item Update $\hat{\lambda}_k = s_1$, $\hat{\bm A}_k=\vec^{-1}_{m_k, n_k}(\bm u_1)$ and $\hat{\bm B}_k=\vec^{-1}_{M-m_k, N-n_k}(\bm v_1)$.
                    \end{itemize}
                    \item Orthonormalize the components by the Gram-Schmidt process in Figure \ref{fig:gram-schmidt-process}.
                \end{itemize}
                \item Output $\{(\hat\lambda_k, \hat{\bm A}_k, \hat{\bm B}_k)\}_{k=1}^{K}$. 
            \end{itemize}
         }
        \end{boxedminipage}
    \end{center}
\end{figure}

\subsection{Hybrid KoPA with Unknown Configurations}\label{sec:method-unknown}

In this section, we consider the case when the model configuration
${\cal C}=\{(m_k, n_k)\}_{k=1}^K$ is unknown. We use a greedy method similar to
boosting to obtain the approximation by iteratively adding one
Kronecker product at a time, based on the residual matrix obtained from
the previous iteration. Specifically, at iteration $k$, we obtain
\[
\hat{\bm E}^{(k)}=\bm Y-\sum_{i=1}^{k-1}\hat{\lambda}_i\hat{\bm A}_i\otimes
  \hat{\bm B}_i,
  \]
  where $\hat{\lambda}_i$, $\hat{\bm A}_i$ and $\hat{\bm B}_i$ are obtained in
  the previous iterations, starting with $\bm E^{(1)}=\bm Y$. Then we use
the single-term KoPA with unknown configuration proposed in
\cite{cai2019kronecker} to obtain
\[
\min_{\lambda_k,\bm A_k, \bm B_k} \|\bm \hat{\bm E}^{(k)} - \lambda_k\bm A_k\otimes \bm B_k\|_F^2,
\]
where the configuration $(m_k,n_k)$
of $\hat{\bm A}_k$ and $\hat{\bm B}_k$ is obtained
by minimizing
the information criterion
\begin{equation}
  IC_q(m, n) = 2^{M+N}\ln \dfrac{\|\hat{\bm E}^{(k)} - \lambda_k\bm A_k\otimes \bm B_k\|_F^2}{2^{M+N}} + qp,\label{eq:ic}
\end{equation}
where $p = 2^{m+n}+2^{M+N-m-n}$ is the number of parameters of
the single-term model with configuration $(m_k, n_k)$ and $q$ is the penalty coefficient on model complexity. When $q=0$ the information criterion is a monotone function of the mean squared error (MSE). When $q=2$ and $q=(M+N)\ln 2$, the information criterion $IC_q(m, n)$ is same as Akaike information criterion (AIC) \citep{Akaike1998information} and Bayes information criterion (BIC) \citep{schwarz1978estimating}, correspondingly.
As shown in \cite{cai2019kronecker}, in a single-term Kronecker product case, when the signal-to-noise ratio is large enough, minimizing the information criterion $IC_q$ in (\ref{eq:ic}) produces consistent estimators of
the true configuration.

The procedure is repeated until a stopping
criterion is reached.
The algorithm is depicted in Figure \ref{fig:iterative}.



\noindent
{\bf Remark:}
The iterative algorithm in Figure \ref{fig:iterative} is a greedy algorithm,
which does not guarantee a global optimal in all configuration combinations.
The output of the algorithm satisfies Assumption \ref{assump:norm-1} but does
not satisfy Assumption \ref{assump:orthogonal}. However, searching the configuration space $\{(m_k, n_k)\}_{k=1}^K$ in a greedy and additive way requires less computational power. It is possible that, given the configuration
$\hat{{\cal C}}=\{(\hat{m}_k,\hat{n}_k), k=1,\ldots, \hat{K}\}$ obtained
in the greed algorithm, a refinement step can be engaged using the algorithm
proposed in Section 3.1. If more computational resources are available,
the refinement can be done at the end of each iteration $k$ based on
$\hat{{\cal C}}_k=\{(\hat{m}_i,\hat{n}_i), i=1,\ldots, k\}$ obtained
to obtain better partial residual matrix
$\hat{\bm E}^{(k)}$ so the configuration determination in future iterations
are more accurate.

\noindent
{\bf Remark:} The stopping criterion can be selected according to the objective
of the study. For denoising applications, one may specify the desired level of
proportion of the total variation explained by the $h$KoPA to be reached.
Similarly a scree plot approach can be used. It is also possible to minimize an overall information criterion in the form
of
\begin{equation}
  IC_q({\cal C}) = 2^{M+N}\ln \dfrac{\|\hat{\bm Y} - \sum_{k=1}^K \lambda_k\bm A_k\otimes \bm B_k\|_F^2}{2^{M+N}} + qp,
\end{equation}
where $p = \sum_{k=1}^K2^{m_k+n_k}+2^{M+N-m_k-n_k}$ is the total
number of parameters with the configuration ${\cal C}$.
For image compression applications,
the information criterion can be replaced with the
ratio of explained variation to
the total storage size. In Section \ref{sec:example-lenna}, we will illustrate a practical stopping criterion using random matrix theory.

\noindent
{\bf Remark:} Compared with the backfitting algorithm for known configuration problems, the iterative algorithm in Figure \ref{fig:iterative} focuses more on the configuration determination in an iterative and additive way. When the configurations are known exactly, the backfitting algorithm in Figure \ref{fig:als} gives more accurate estimates of the components.

\begin{figure}[!htb]
    \begin{center}
        \begin{boxedminipage}{6.5in}
        \caption{Iterative Algorithm for $h$KoPA Estimation}
        \label{fig:iterative}
        \it{
            \begin{itemize}
                \item Set $\hat{\bm E}^{(1)}=\bm Y$.
                \item For $k=1,\dots, K$:
                \begin{itemize}
                    \item For all $(m, n)\in \mathcal C:=\{0,\dots, M\}\times\{0,\dots, N\}/\{(0, 0), (M, N)\}$:
                    \begin{itemize}
                        \item Do SVD for $\mathcal R_{m,n}[\hat{\bm E}^{(k)}]$ such that
                        $$\mathcal R_{m,n}[\hat{\bm E}^{(k)}]=\sum_{j=1}^Js_j\bm u_j\bm v_j^T.$$
                        \item Set $\hat\lambda_k^{(m, n)}=s_1$, $\hat{\bm A}_k^{(m, n)}=\vec^{-1}_{m, n}(\bm u_1)$ and $\hat{\bm B}_k^{(m, n)}=\vec^{-1}_{M-m, N-n}(\bm v_1)$.
                        \item Obtain $\hat{\bm S}_k^{(m, n)} = \hat\lambda_k^{(m, n)}\hat{\bm A}_k^{(m, n)}\otimes\hat{\bm B}_k^{(m, n)}$.
                    \end{itemize}
                    \item Obtain $(\hat m_k,\hat n_k)$ through
                    $$\hat m_k, \hat n_k = \argmin_{(m, n)\in\mathcal C}\ 2^{M+N}\ln \dfrac{\|\hat{\bm E}^{(k)} - \hat{\bm S}_k^{(m, n)}\|_F^2}{2^{M+N}}+qp.$$
                    \item Set $\hat\lambda_k = \hat\lambda_k^{(\hat m_k,\hat n_k)}$, $\hat{\bm A}_k = \hat{\bm A}_k^{(\hat m_k,\hat n_k)}$ and $\hat{\bm B}_k = \hat{\bm B}_k^{(\hat m_k,\hat n_k)}$.
                    \item Break if a stopping criterion is met.
                    \item Calculate $\hat{\bm E}^{(k+1)} = \hat{\bm E}^{(k)} - \hat{\bm S}_k^{(\hat m_k, \hat n_k)}$.
                \end{itemize}
              \item Return $\{(\hat\lambda_k, \hat{\bm A}_k, \hat{\bm B}_k)\}_{k=1}^{\hat K}$, where $\hat K$ is the number of terms determined by the stopping criterion.
            \end{itemize}
         }
        \end{boxedminipage}
    \end{center}
\end{figure}

\section{Empirical Examples}\label{sec:example}
\subsection{Simulation}\label{sec:example-simulation}
In this simulation, we exam the performance of the least squares backfitting
algorithm in Figure \ref{fig:als} for a two-term Kronecker product model and
determine
the factors that may affect the accuracy and convergence speed of the
algorithm.

Specifically, we simulate the data $\bm Y$ according to
$$\bm Y = \lambda_1\bm A_1\otimes \bm B_1 + \lambda_2\bm A_2\otimes \bm B_2 + \sigma \bm E,$$
where $\lambda_1=\lambda_2=1$ and $\bm A_k$, $\bm B_k$ ($k=1,2$)
satisfy Assumption \ref{assump:norm-1}.
Here we assume $\bm A_1$ and $\bm A_2$ are linearly independent to ensure identifiability Assumption \ref{assump:orthogonal}, as discussed in Section 2.
If they are not linearly independent, we
can always apply Gram-Schmidt process to reformulate the model so
$\bm A_1$ and $\bm A_2$ are linearly independent. One of the objectives of
the simulation study is to see the impact of linear dependence of
$\bm B_1$ and $\bm B_2$, once we control the
linear dependence of $\bm A_1$ and $\bm A_2$.
The shape parameters for
this simulation study is set to
$$M=N=9,\quad (m_1, n_1) = (4, 4),\quad (m_2, n_2)=(5, 5).$$

To simulate the component matrices, we first generate $\tilde{\bm A}_1\in\mathbb R^{2^{m_1}\times 2^{n_1}}$ and $\tilde{\bm A}_2\in\mathbb R^{2^{m_2}\times 2^{n_2}}$ by normalizing corresponding i.i.d standard Gaussian random matrices.
$\tilde{\bm A}_1$ and $\tilde{\bm A}_2$ are then orthogonalized in the sense that $\|\tilde{\bm A}_1\otimes \bm C - \tilde{\bm A}_2\|_F^2$ is minimized at $\bm C = 0$. Orthonormal $\tilde {\bm B}_1$ and $\tilde {\bm B}_2$ are generated in a similar way. We set
$$\bm A_1 = \tilde{\bm A}_1,\ \bm A_2 = \tilde{\bm A}_2,\ \bm B_1 = \dfrac{\tilde {\bm B}_1 + \alpha \bm 1\otimes\tilde {\bm B}_2}{\sqrt{1 + \alpha^2 2^{m_2+n_2-m_1-n_1}}},\ \bm B_2 = \tilde {\bm B}_2,$$
where $\bm 1$ is a $2^{m_2-m_1}\times 2^{n_2-n_1}$ matrix of ones and $\alpha$ controls the linear dependency between $\bm B_1$ and $\bm B_2$.

Note that when $\alpha = 0$, $\bm B_1$ and $\bm B_2$ are linearly independent. When $\alpha \rightarrow\infty$, $\bm B_1\propto \bm 1\otimes \bm B_2$
and the model can be represented with a single term Kronecker product.

In this simulation we consider $\alpha\in
\{ 0, 0.5, 1.0, 1.5, 2.0\}$  and
$\sigma_0:=2^{(M+N)/2}\sigma \in \{0, 0.5, 1.0, 1.5, 2.0\}$.
The benchmark setting is $\alpha = 0.5$ and $\sigma_0=1$, under which the signal-to-noise ratio is $(\lambda_1^2+\lambda_2^2)/\sigma_0^2=2$ and $50\%$ of the variation in $\bm B_1$ can be explained by $\bm B_2$.

\begin{figure}[!htb]
    \centering
    \includegraphics[scale=0.6]{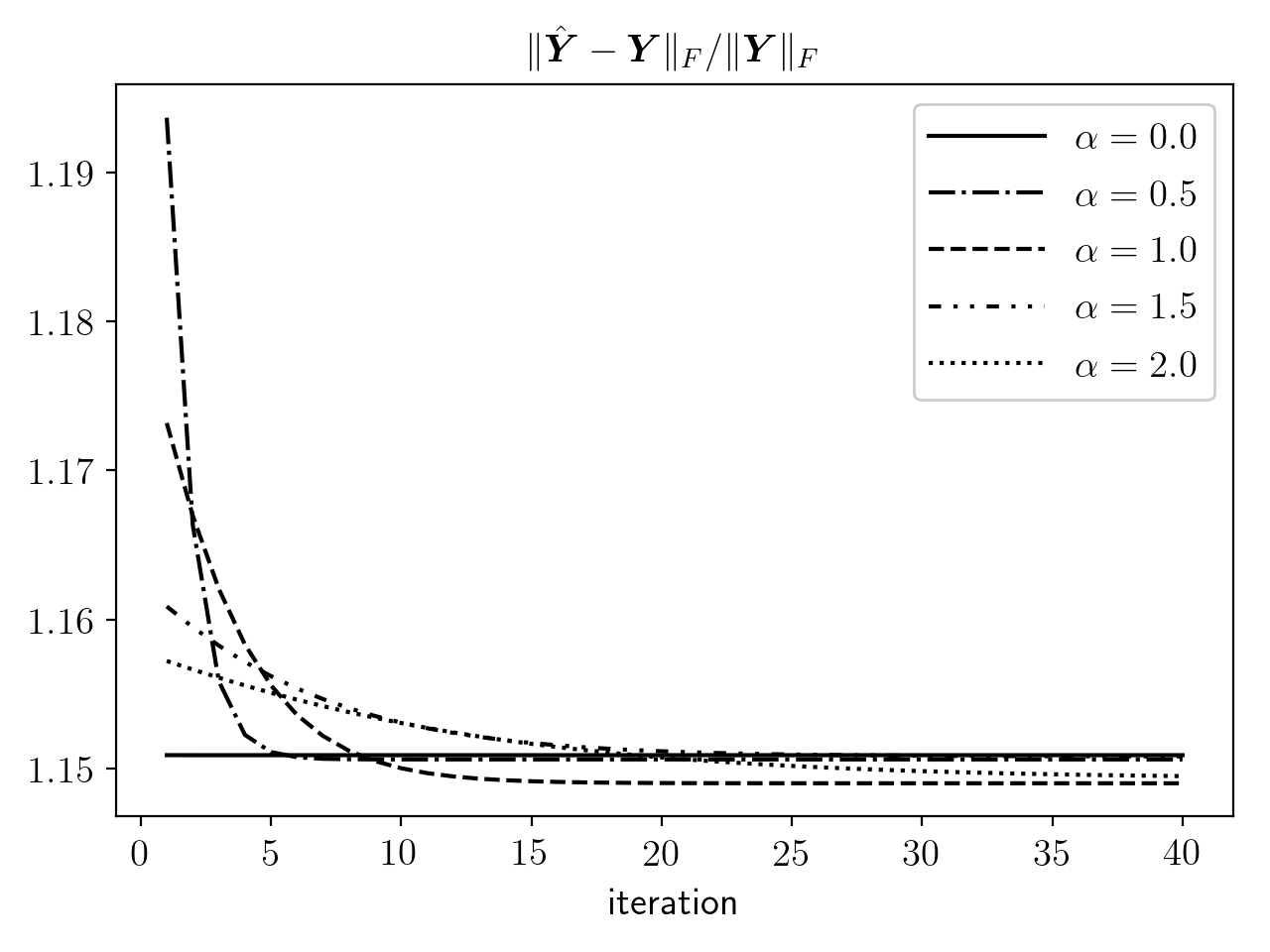}
    \caption{Fitting error against number of iterations for different $\alpha$ values, when $\sigma_0=1$}
    \label{fig:y-vs-alpha}
\end{figure}

We first examine the effect of linear dependency of
$\bm B_1$ and $\bm B_2$, controlled by $\alpha$.
Hence we fix $\sigma_0=1$ and check the performance of the
backfitting algorithm under different values of $\alpha$, using the known configurations. Figure \ref{fig:y-vs-alpha} shows the relative error $\hat{\bm Y}$ for the first 40 iterations for the five different values of $\alpha$. At $\sigma_0=1$, a perfect fit
is expected to have an error of
$$\dfrac{E[\|\sigma\bm E\|_F]}{E[\|\bm Y\|_F]}
= \dfrac{\sigma_0}{\sqrt{\lambda_1^2+\lambda_2^2+\sigma_0^2}}\approx 0.577.$$
under the simulation setting. 
It is seen from Figure \ref{fig:y-vs-alpha}
that the estimates tend to overfit as the final relative errors are all smaller
than the expected value. This is due to the fact that
the observed error term $\bm E$ is not orthogonal to the observed signal.
However, the less $\bm B_1$ and $\bm B_2$ are linearly dependent, the less
the model is overfitted.


\begin{figure}[!htpb]
    \centering
    \includegraphics[width=0.45\textwidth]{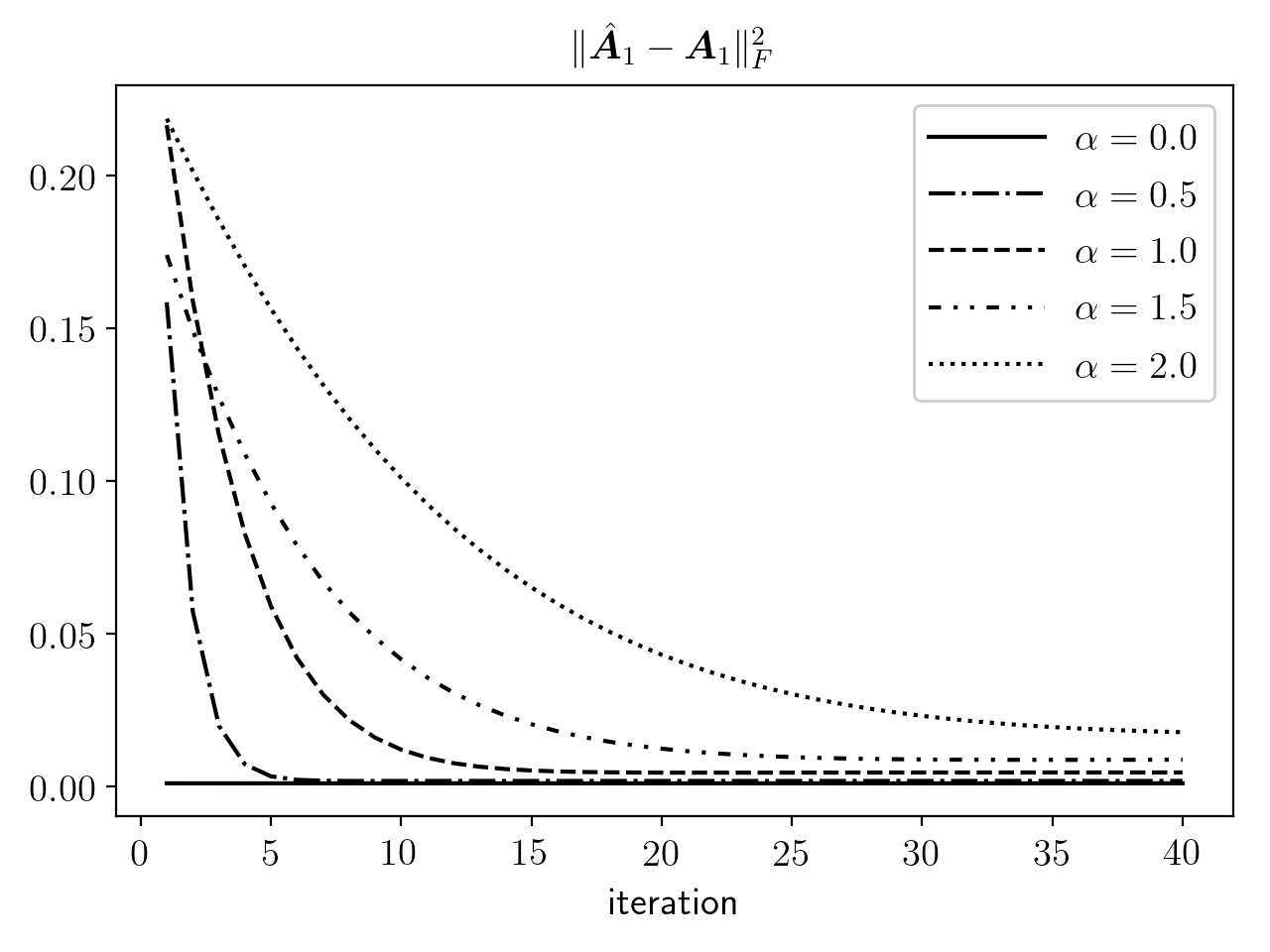}
    \includegraphics[width=0.45\textwidth]{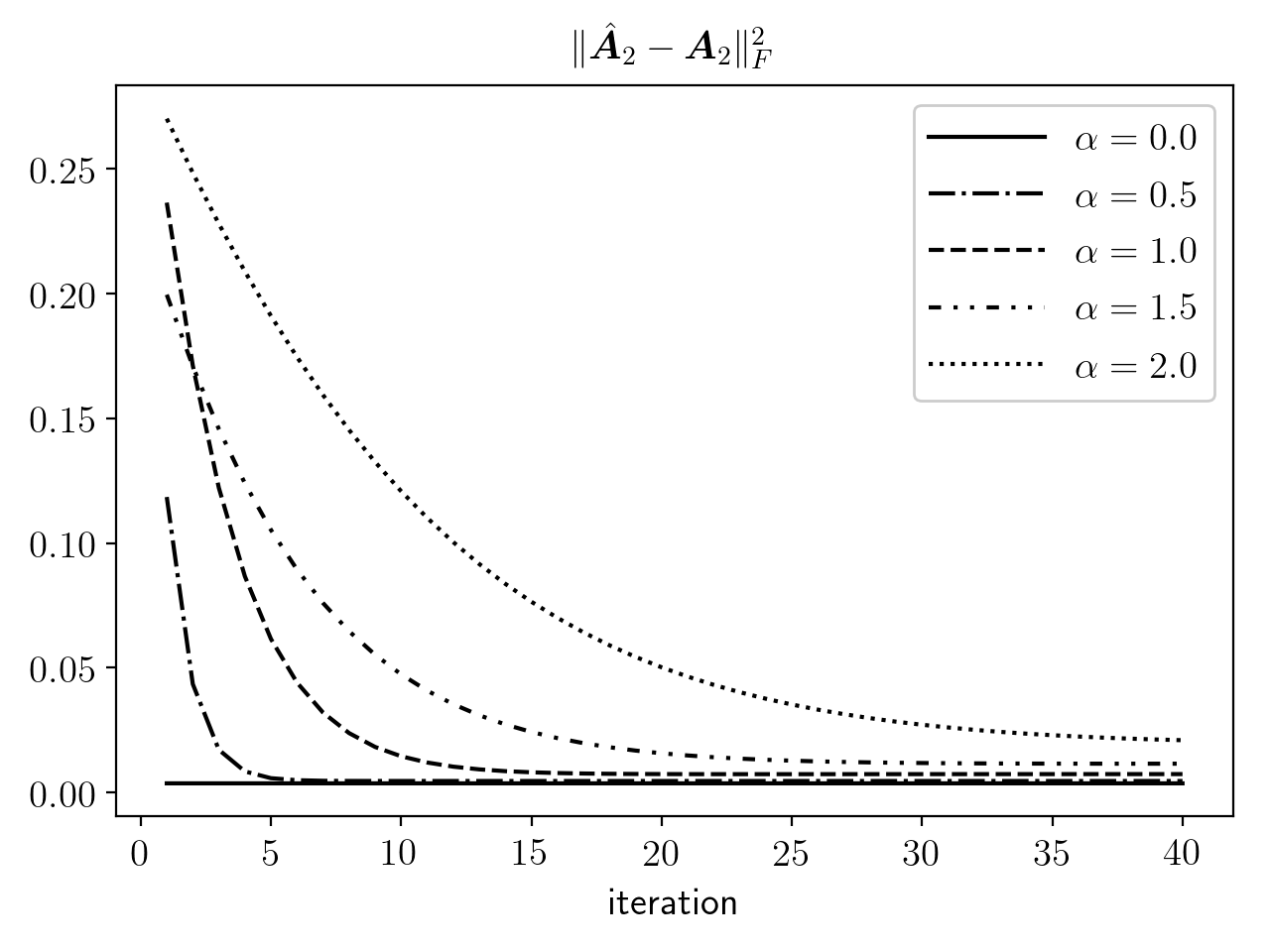}\\
    \includegraphics[width=0.45\textwidth]{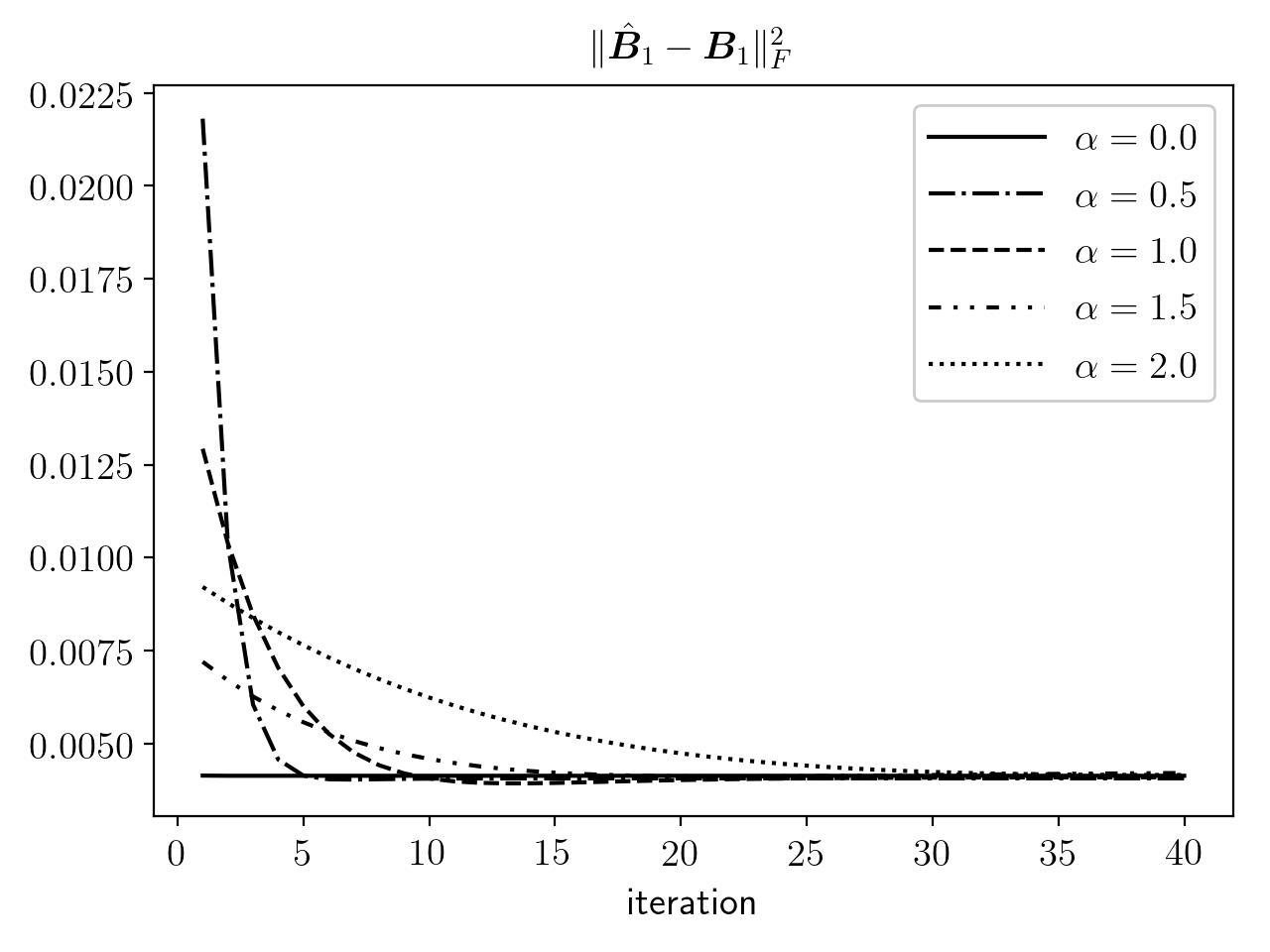}
    \includegraphics[width=0.45\textwidth]{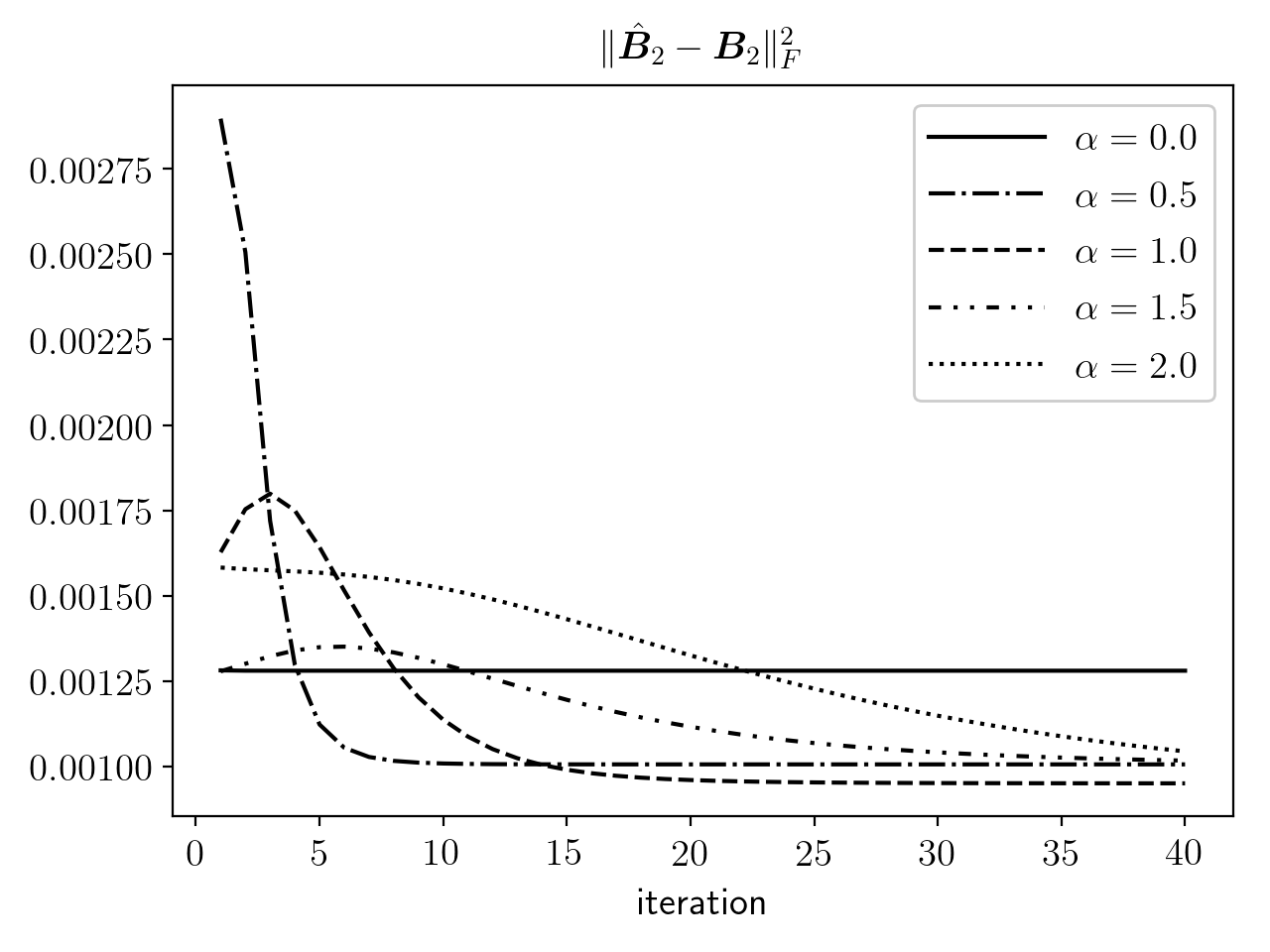}\\
    \includegraphics[width=0.45\textwidth]{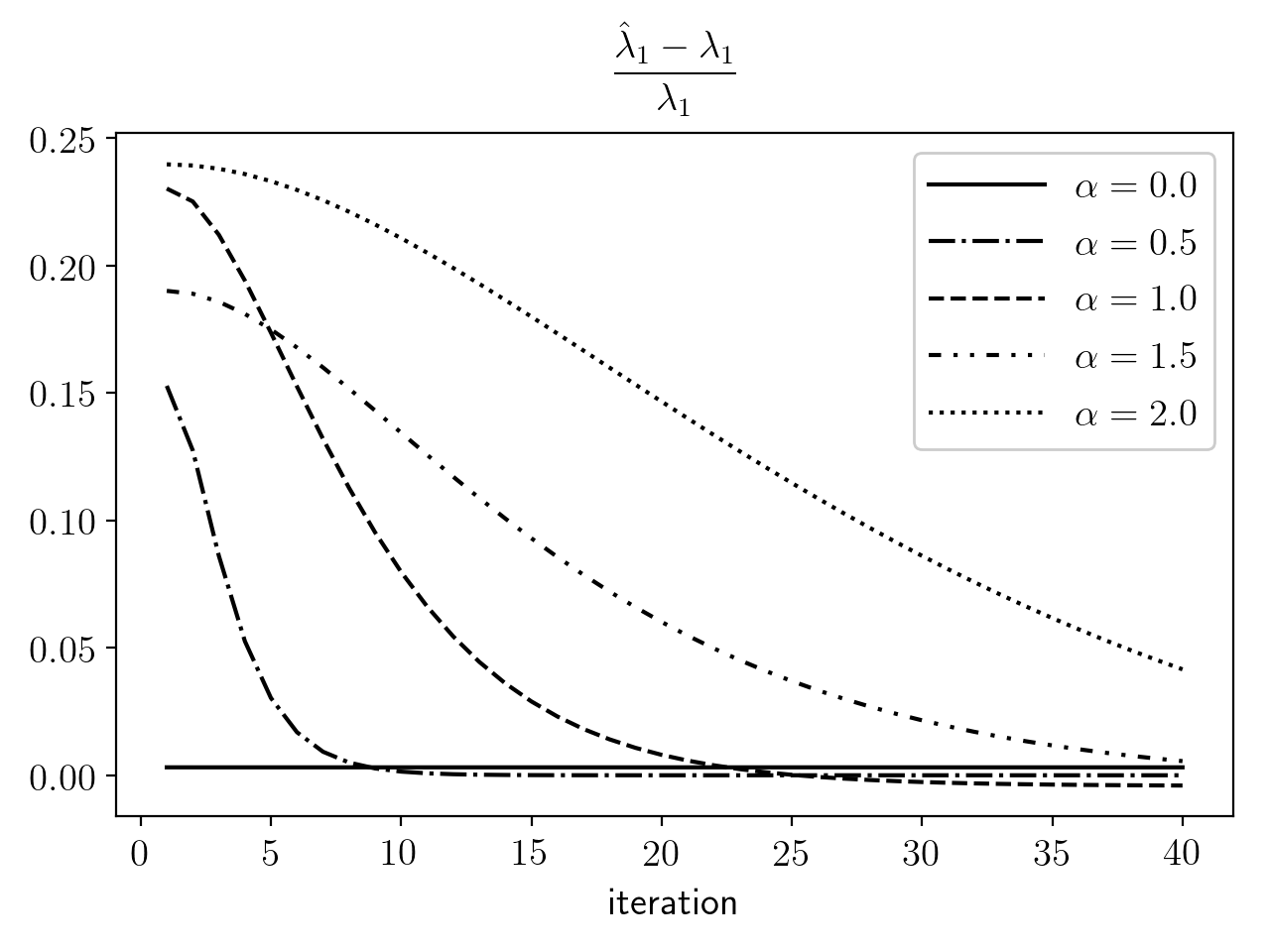}
    \includegraphics[width=0.45\textwidth]{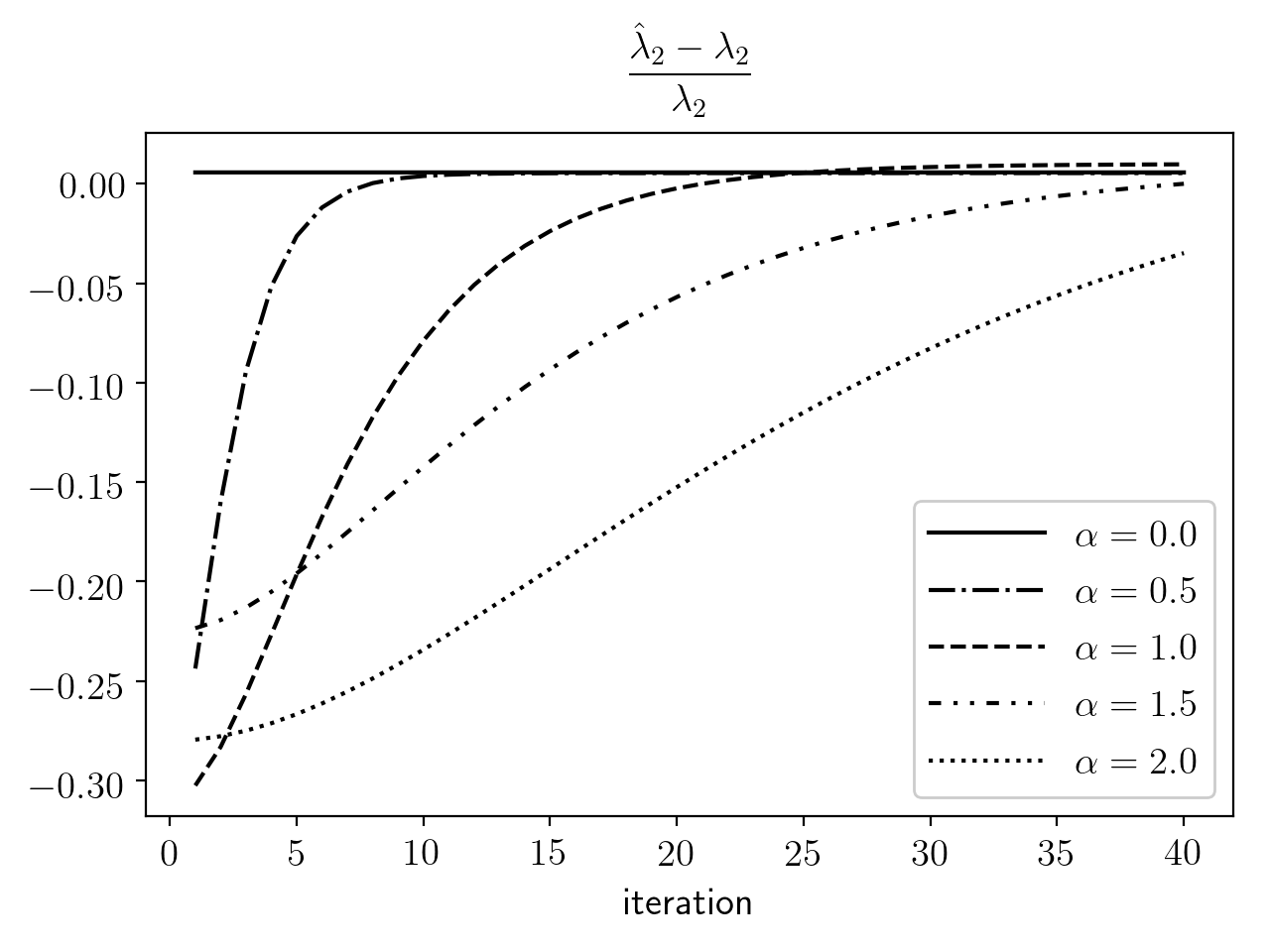}
    \caption{Errors in components against number of iterations for different $\alpha$ values.}
    \label{fig:error-vs-alpha}
\end{figure}

By comparing the convergence speed of different $\alpha$ values, we notice that larger value of $\alpha$, corresponding to higher linear dependency between
$\bm B_1$ and $\bm B_2$, results in a slower convergence rate. When
$\bm B_1$ and $\bm B_2$ are linearly independent
($\alpha= 0$), only one iteration is needed.

The errors of the estimation of the coefficients $\lambda_1$, $\lambda_2$,
$\bm A_1$, $\bm A_2$, $\bm B_1$ and $\bm B_2$ plotted against the number
of iterations are shown in Figure~\ref{fig:error-vs-alpha}.
For all the components, higher values of $\alpha$
result in slower convergence rates and less accurate final results.

Notice that when $\alpha > 0$, the four matrices involved in the two-term model are not symmetric under the identifiability conditions using the iterative backfitting algorithm. On the one hand, $\bm A_1$ and $\bm A_2$ are enforced to be orthogonal, while $\bm B_1$ and $\bm B_2$ are not. On the other hand, the iterative backfitting algorithm fits $\bm A_1$ and $\bm B_1$ first, resulting in an overfitting for the first Kronecker product in the first iteration. 
Due to the asymmetry, the performance of estimating $\bm B_1$ and $\bm B_2$ is better than that of estimating $\bm A_1$ and $\bm A_2$ when $\bm B_1$ and $\bm B_2$ have linear dependency. A strong correlation between the errors in $\bm A_1$ and $\bm A_2$ is observed.
We also note that the convergence rate for
both $\lambda_1$ and $\lambda_2$ are slower than that for the matrices,
especially for the high linear dependence cases.

\begin{figure}[!htb]
    \centering
    \includegraphics[scale=0.6]{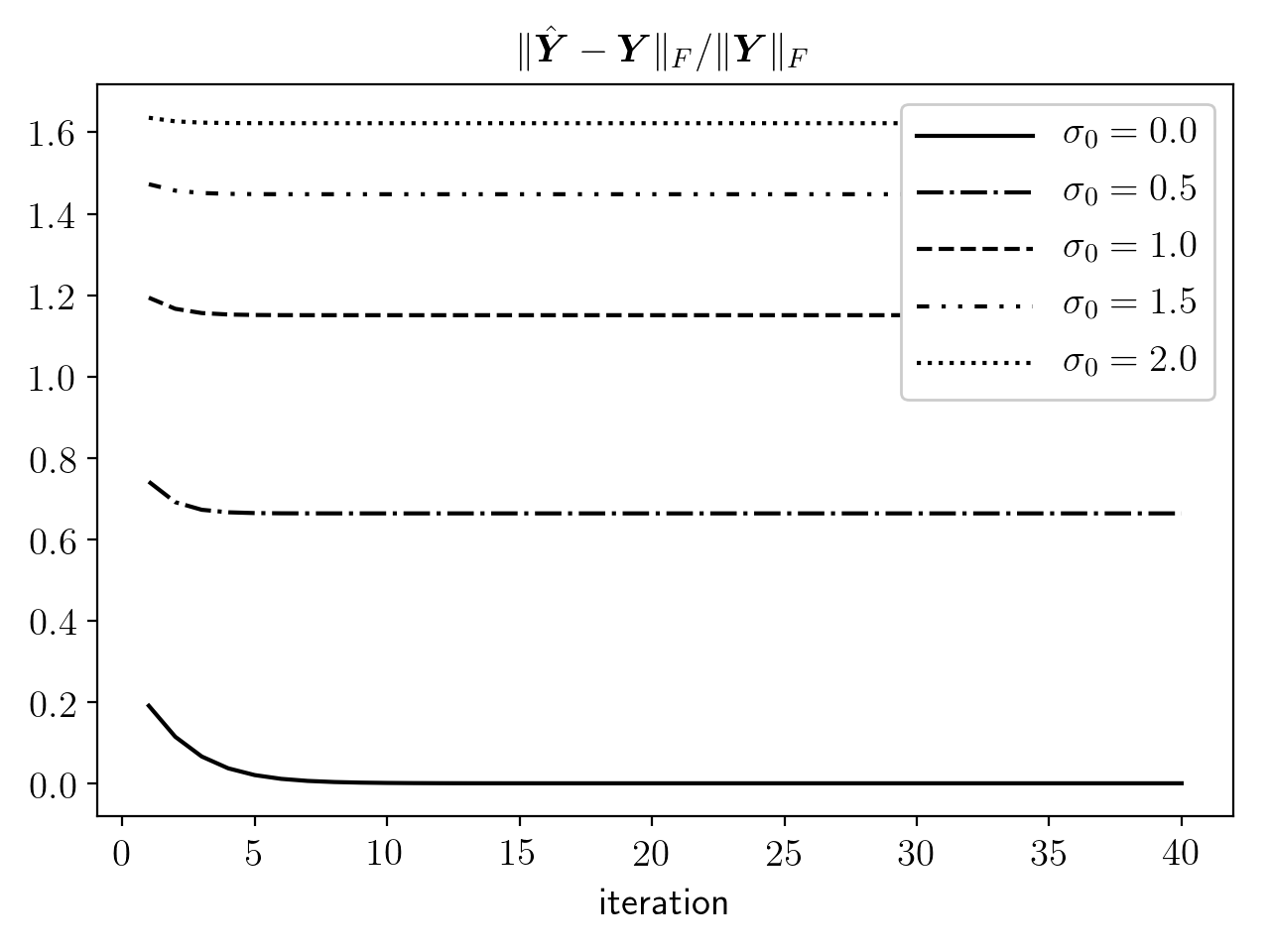}
    \caption{Fitting error against number of iterations for different $\sigma_0$ values}
    \label{fig:y-vs-sigma}
\end{figure}

Next, we examine the effect of the noise level $\sigma_0$. We fix $\alpha = 0.5$ and consider five different values of the noise $\sigma_0$. The error in estimating
$\bm Y$ is reported in Figure \ref{fig:y-vs-sigma}. It is seen that
higher noise level $\sigma_0$ results in larger errors, as expected.
A small difference in the convergence speed is observed as well.
The algorithm
converges faster when the noise level is high, but it is not as sensitive
as that in the change of linear dependence level $\alpha$.

\begin{figure}[!htpb]
    \centering
    \includegraphics[width=0.45\textwidth]{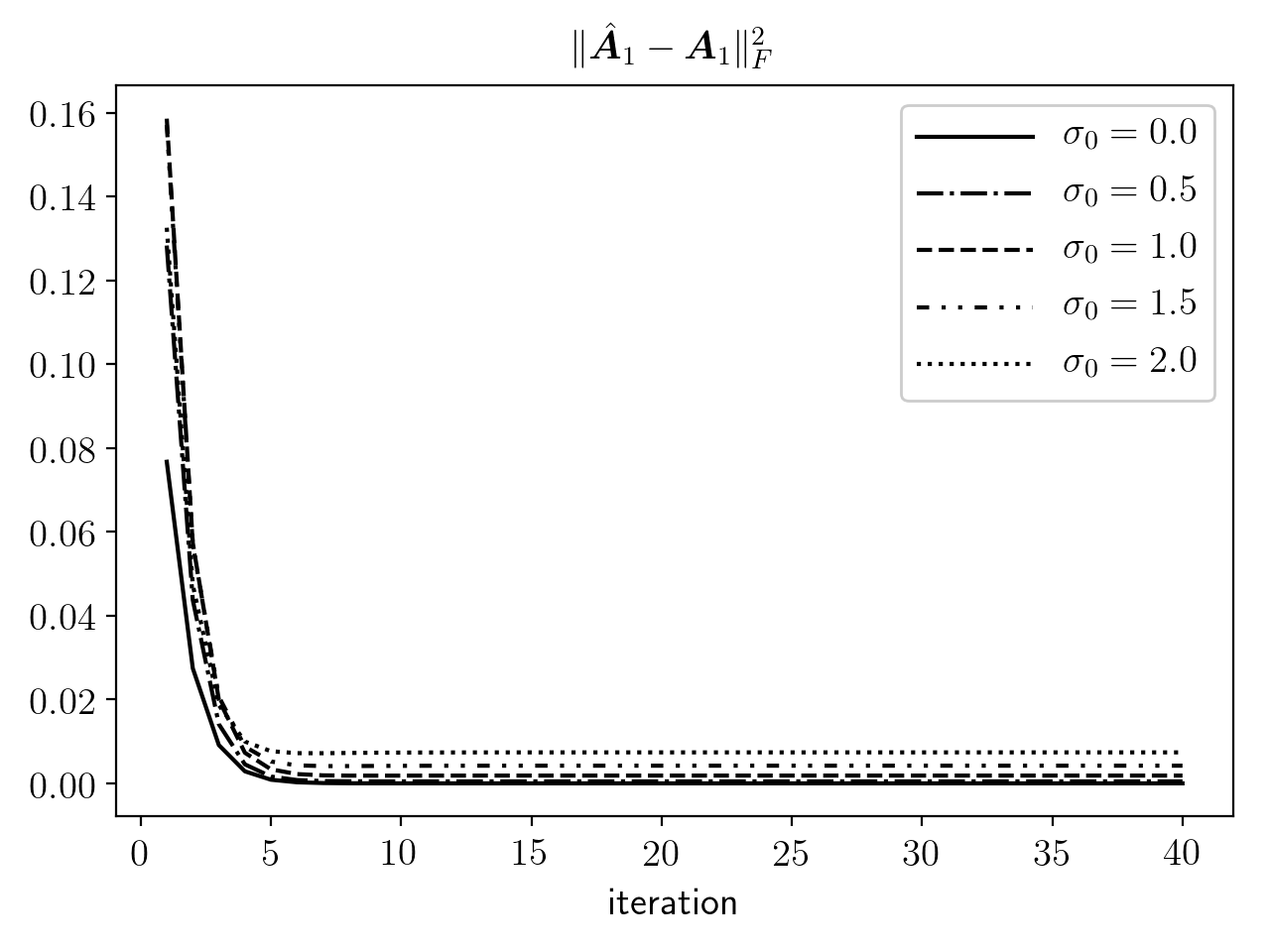}
    \includegraphics[width=0.45\textwidth]{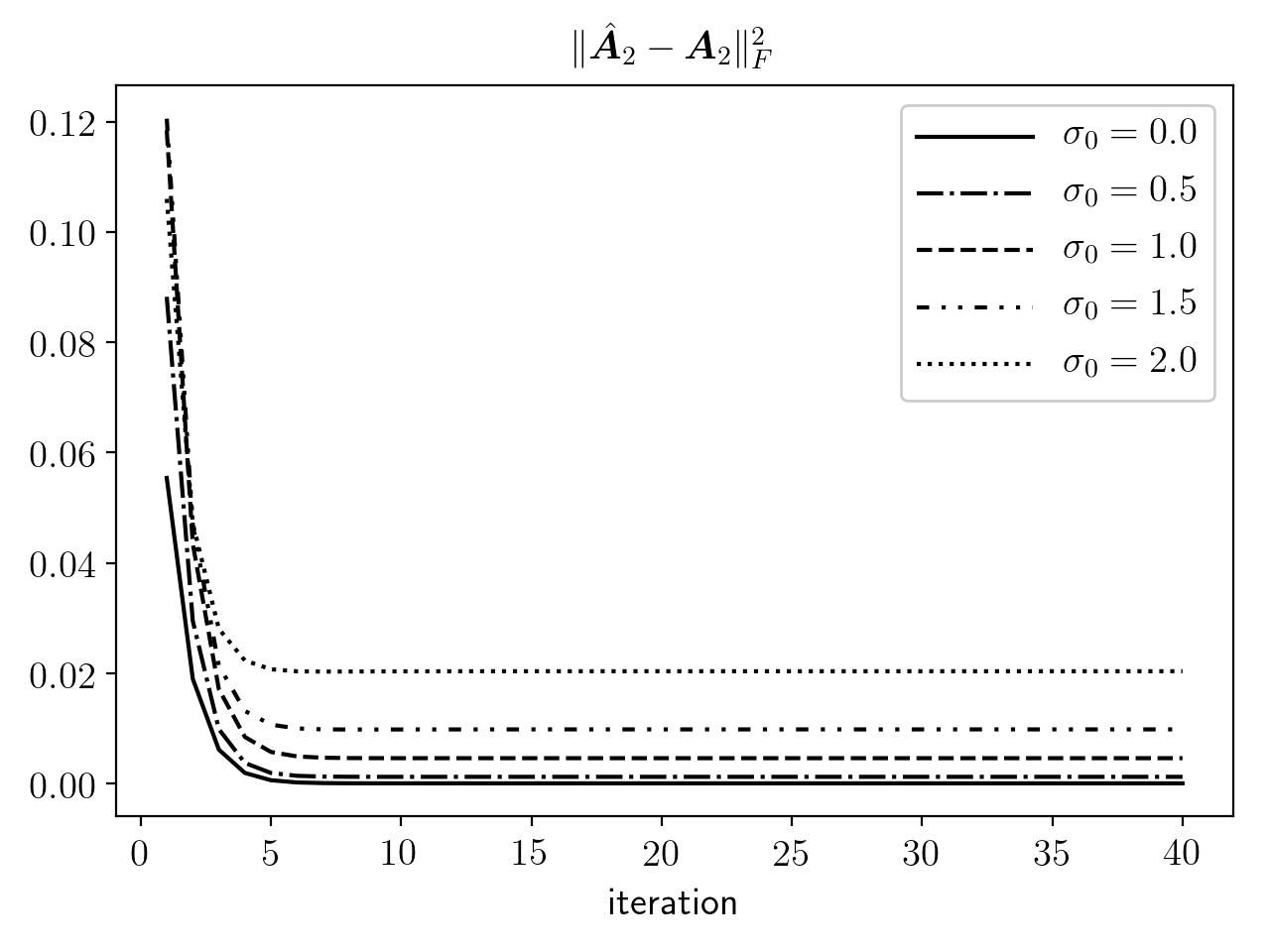}\\
    \includegraphics[width=0.45\textwidth]{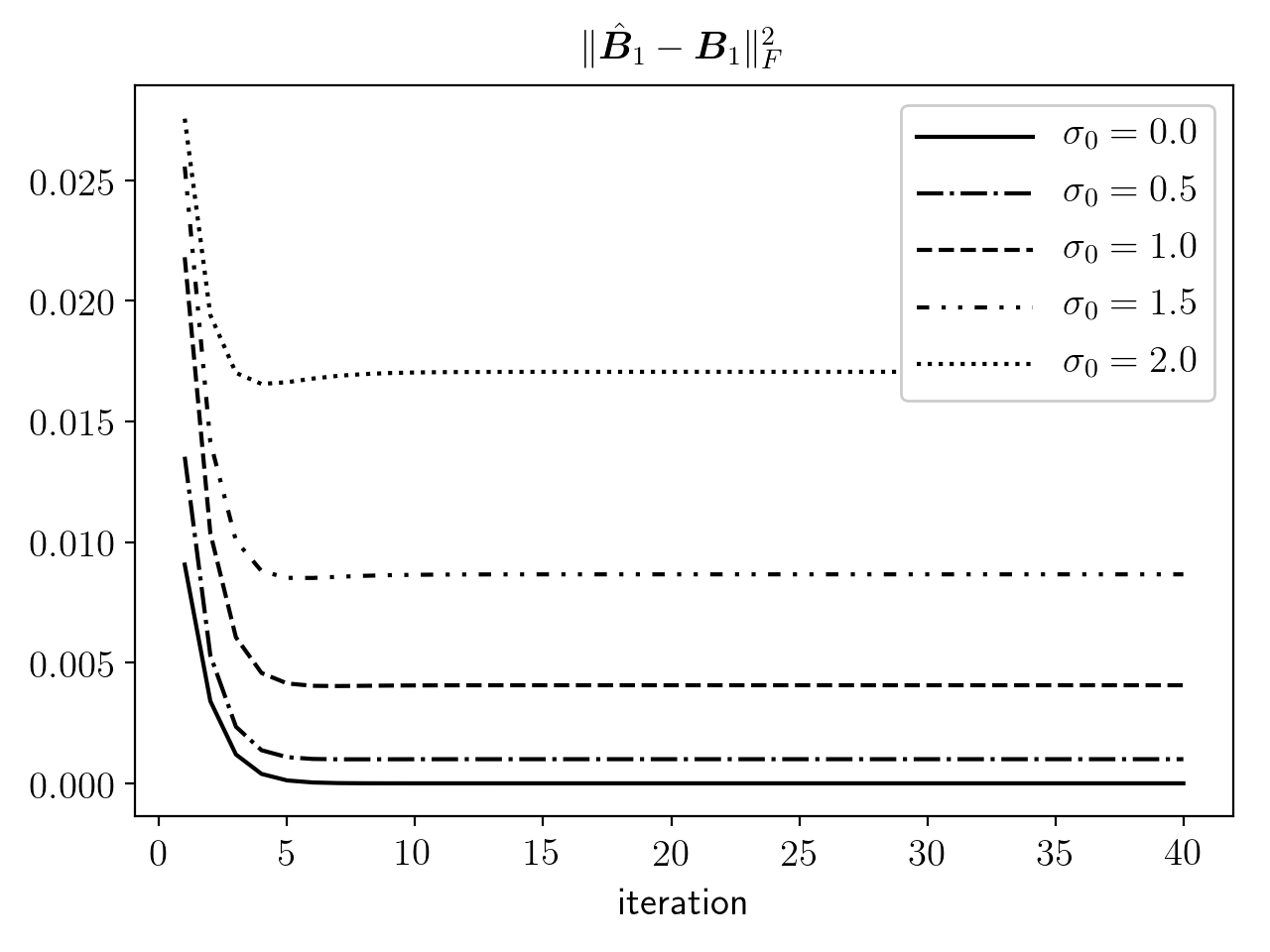}
    \includegraphics[width=0.45\textwidth]{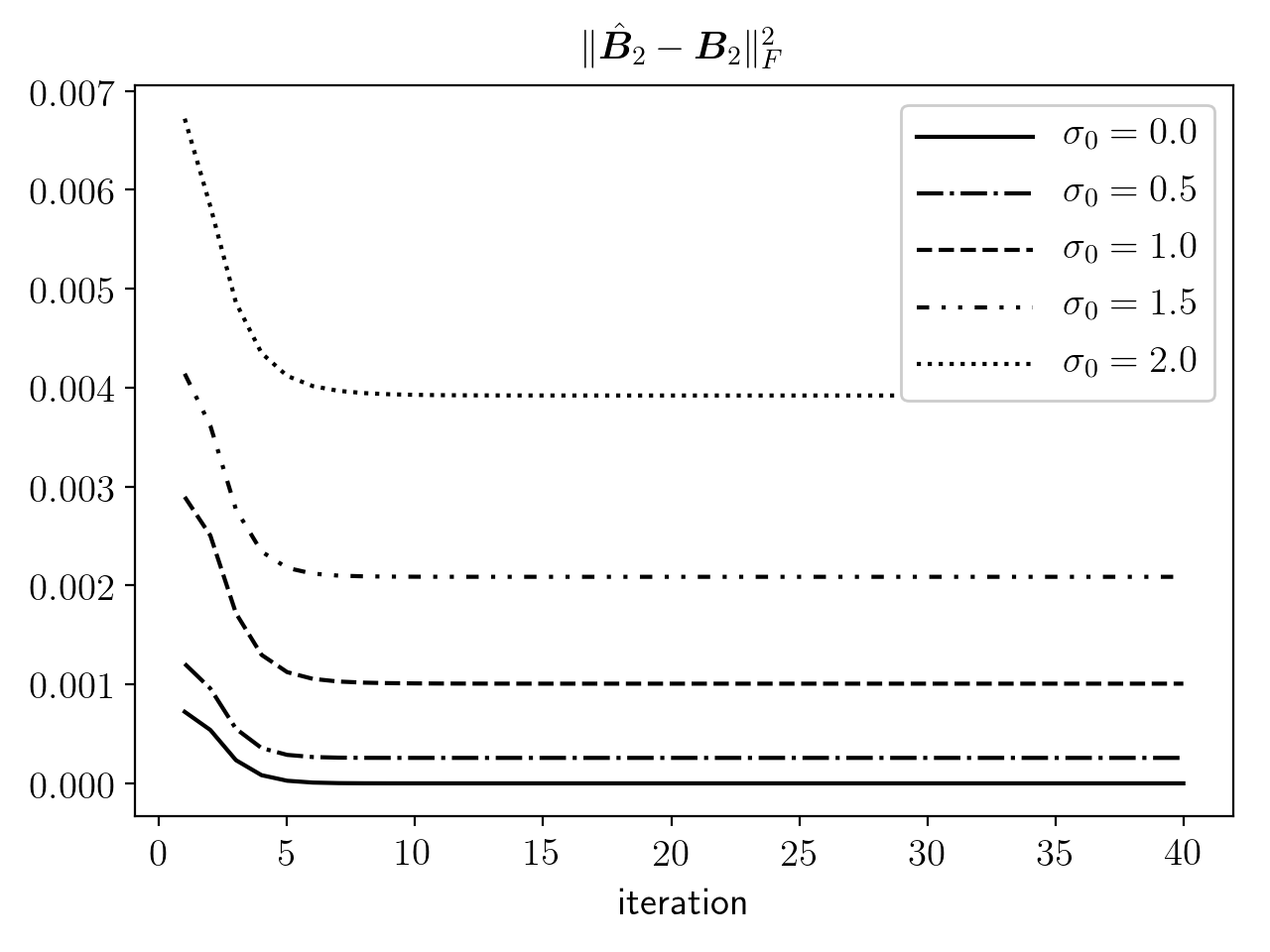}\\
    \includegraphics[width=0.45\textwidth]{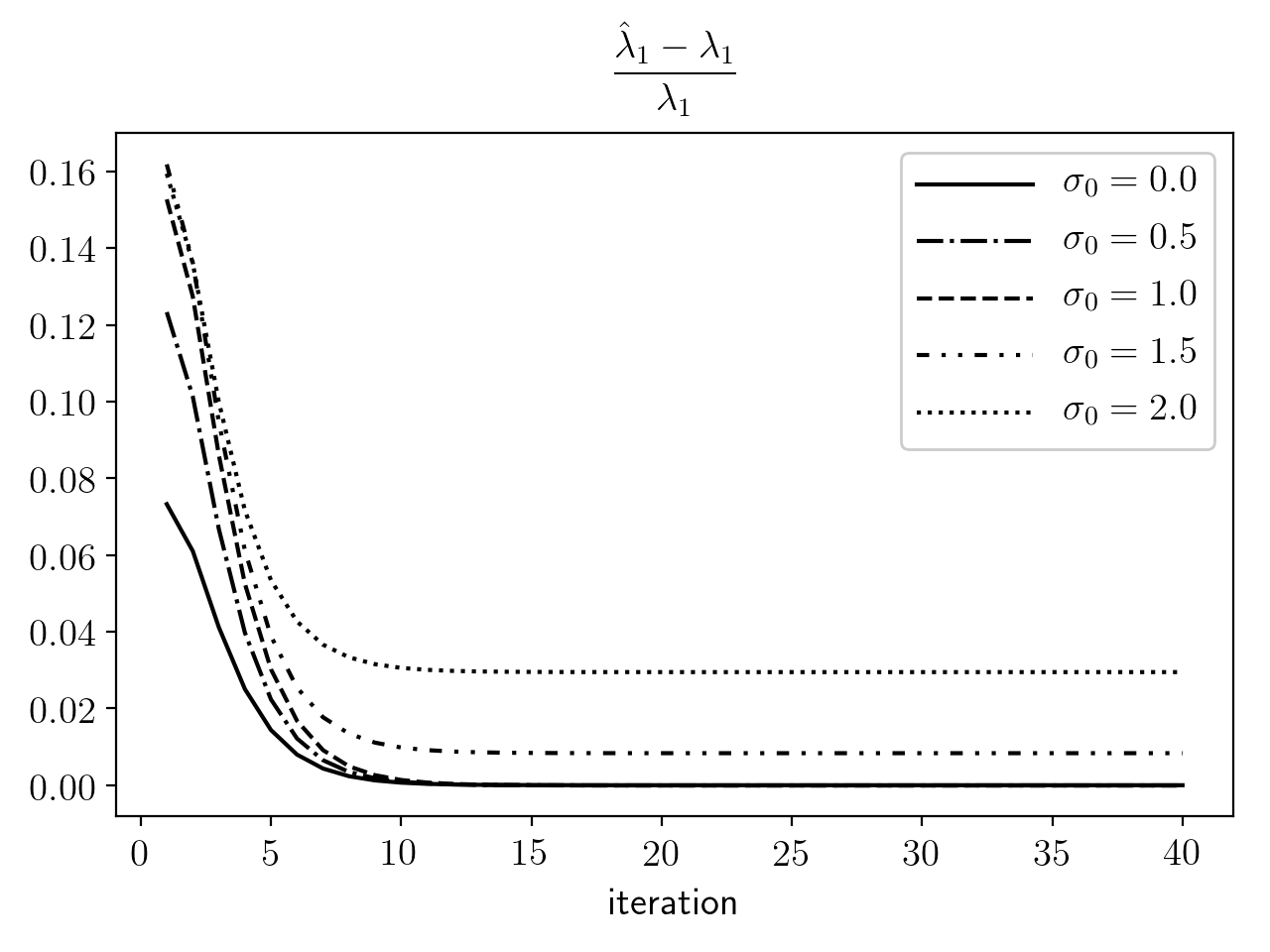}
    \includegraphics[width=0.45\textwidth]{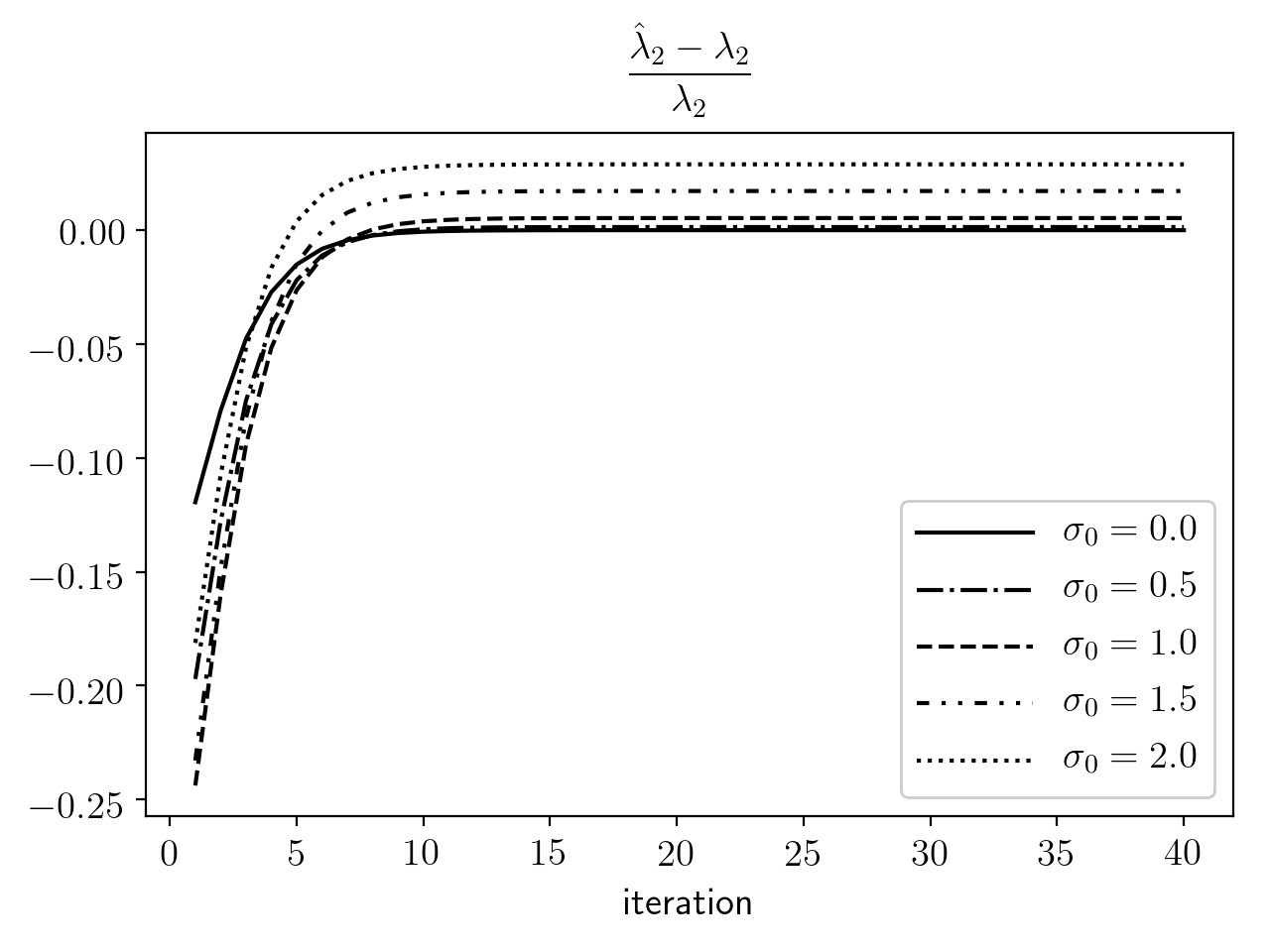}
    \caption{Errors in components against number of iterations for different $\sigma_0$ values.}
    \label{fig:error-vs-sigma}
\end{figure}

Errors for estimating the different components in the model are plotted in Figure \ref{fig:error-vs-sigma} for different noise levels. The difference in convergence rates is less obvious. 
We also observe that the performance for estimating the smaller component matrices $\bm A_1$ and $\bm B_2$ is better than that for the larger matrices $\bm A_2$ and $\bm B_1$.



\subsection{Real image example}\label{sec:example-lenna}
\begin{figure}[!htb]
    \centering
    \includegraphics[scale=0.5]{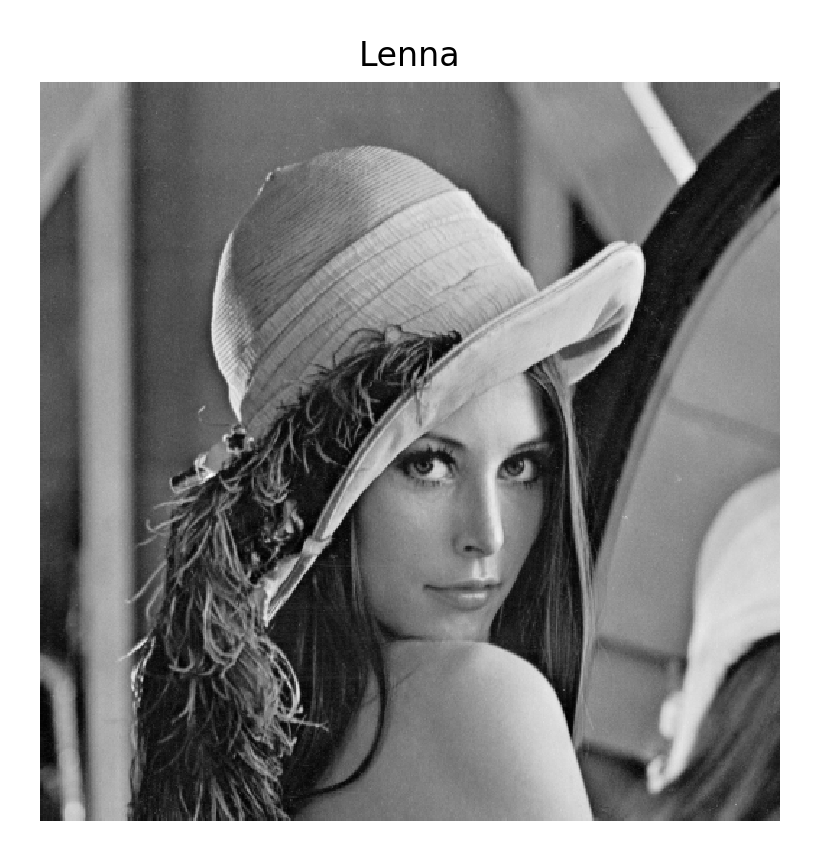}
    \caption{Lenna's image}
    \label{fig:lenna}
\end{figure}

\begin{figure}[!htb]
    \centering
    \includegraphics[scale=0.45]{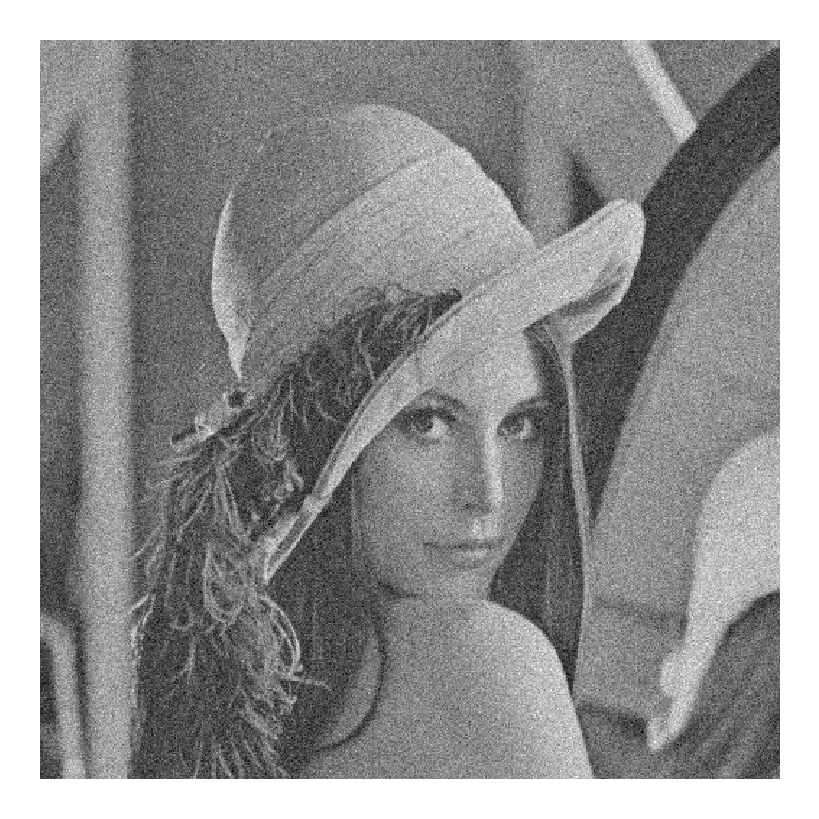}
    \includegraphics[scale=0.45]{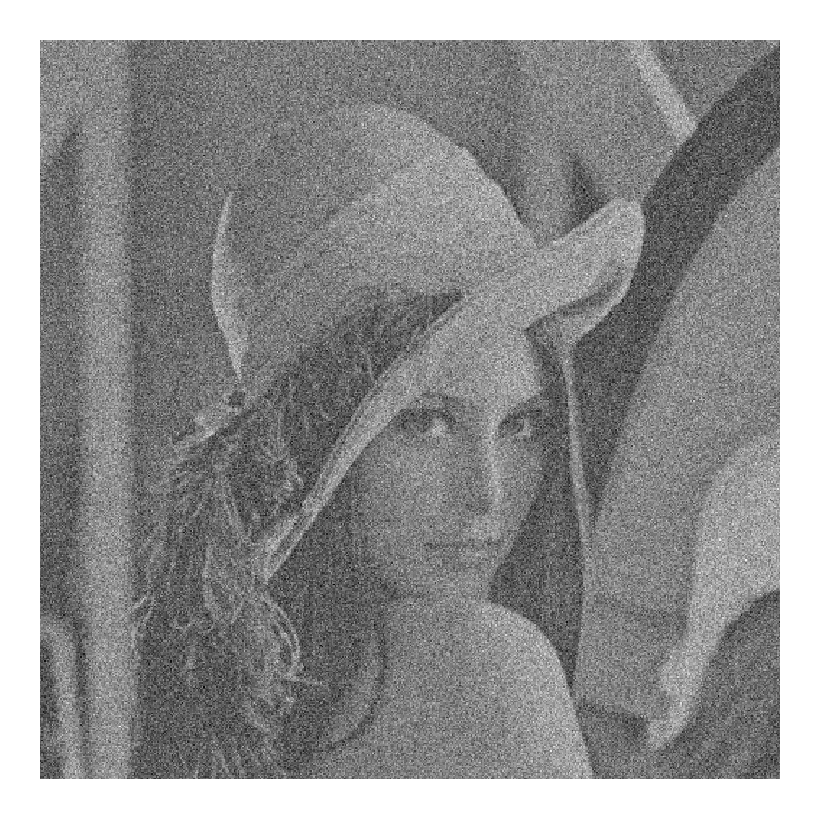}
    \includegraphics[scale=0.45]{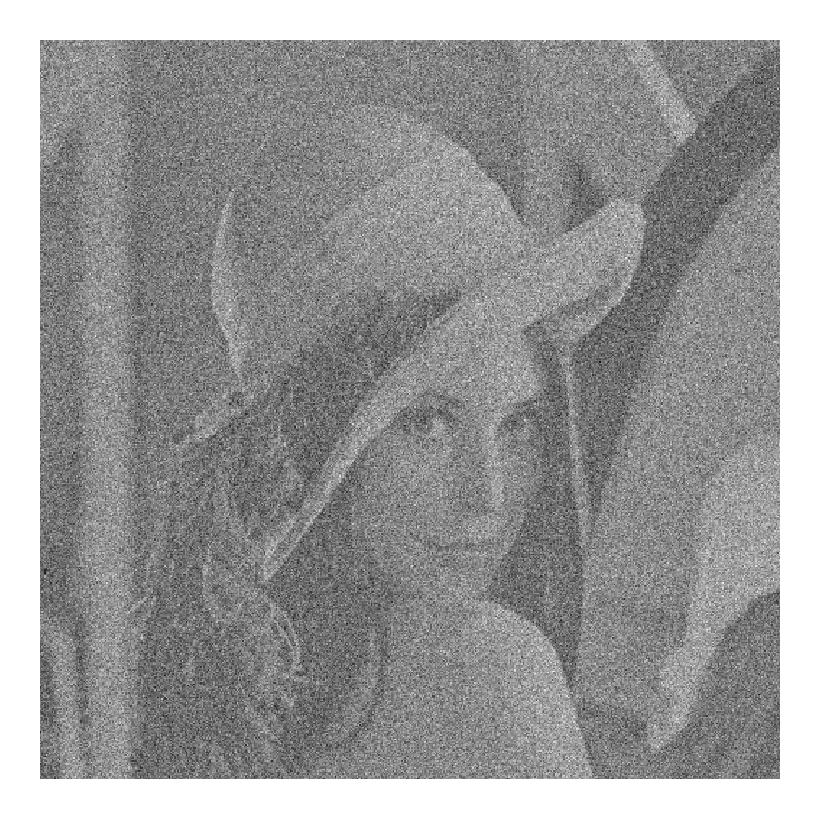}
    \caption{Lenna's image with three noise levels $\sigma \in \{0.1,0.2,0.3\}$}
    \label{fig:lenna_blurred}
\end{figure}

In this exercise, we apply the $h$KoPA to analyze the image of Lenna, which has been used widely as a benchmark image in image processing studies. The Lenna's image shown in Figure \ref{fig:lenna} is a gray-scaled $512\times 512$ picture, which is represented by a $512\times 512$ ($M=N=9$)
real matrix $\bm Y$. The elements of $\bm Y$ are real numbers between 0 and 1, where $0$ represents black and $1$ represents white. Besides the original image, in this example we also consider some artificially
blurred images using
$$\bm Y_\sigma = \bm Y + \sigma \bm E,$$
where $\bm E$ is a matrix of i.i.d. standard Gaussian random variables and $\sigma$ denotes the noise level.
Here we consider three noise levels $\sigma\in \{0.1, 0.2, 0.3\}$. Note that
the original image scale is $[0,1]$. Hence the image with noise level
$\sigma=0.3$ is
considered to be heavily blurred. The blurred images are shown in Figure \ref{fig:lenna_blurred}.

For this real image, the configurations in the $h$KoPA model (\ref{eq:mtkpm})
is unknown. Therefore, we adopt the iterative greedy algorithm proposed in Section \ref{sec:method-unknown}, where the configuration in each iteration is
determined by BIC. For each $\bm Y_\sigma$, we consider to fit the image with at most 20 Kronecker product terms. The configurations selected $(\hat m_k, \hat n_k)$, the fitted $\hat\lambda_k$ and the cumulated percentage of variance explained (c.p.v.) for the first 10 iterations are reported in Table~\ref{tab:iteration}.
It is seen that for all noise level $\sigma$ values, the first several Kronecker products terms can explain most of the variances of $\bm Y_\sigma$.
With larger $\sigma$, $\lambda_k$ decreases slower as $k$ increases, because the residual matrix $\bm E^{(k)}$ has smaller variance.
We also note that as the number of terms $k$ increases, configurations with smaller number of parameters tend to be selected, with some exceptions. This is partially because the residual matrix $\bm E^{(k)}$ tends to become less complex as higher level complexity tends to be explained by the earlier terms. Note that configuration $(6,7)$ has total $2^{6+7}+2^{3+2}-1=8223$ parameters, while $(4,5)$ has total $2^{4+5}+2^{5+4}-1=1023$ parameters. The proportion of variance of the true image $\bm Y$ is shown in the bottom row in Table \ref{tab:iteration}. An overfitting is observed for $\sigma =0.3$ after the seventh iteration.

In the heavily blurred cases, configurations close to the center such as $(5, 4)$ are more likely to be selected by BIC. These configurations correspond to
more squared $\bm A_k$ and $\bm B_k$ matrices.

\begin{table}[!htb]
    \centering
    \setlength{\tabcolsep}{5pt}
    \begin{tabular}{|c|c|c|c|c|c|c|c|c|c|c|c|c|}
        \hline 
         \multirow{2}{*}{k}& \multicolumn{3}{|c|}{$\sigma = 0.0$}& \multicolumn{3}{|c|}{$\sigma = 0.1$} & \multicolumn{3}{|c|}{$\sigma = 0.2$} & \multicolumn{3}{|c|}{$\sigma = 0.3$} \\ \cline{2 - 13}
         & $(\hat m_k, \hat n_k)$ & $\hat\lambda_k$ & c.p.v.& $(\hat m_k, \hat n_k)$ & $\hat\lambda_k$ & c.p.v.& $(\hat m_k, \hat n_k)$ & $\hat\lambda_k$ & c.p.v. & $(\hat m_k, \hat n_k)$ & $\hat\lambda_k$ & c.p.v.\\
         \hline
         1 & (6, 7) & 95.07  & 91.81 & (5, 6) & 91.21 & 66.76 & (5, 6) & 91.80 & 41.40 & (4, 6) & 88.28 & 23.21\\
         2 & (6, 6) & 14.21 & 93.86 & (5, 6) & 21.88 & 70.60 & (3, 6) & 15.42 & 42.57 & (4, 5) & 26.97 & 25.37\\
         3 & (5, 7) & 12.18 & 95.39 & (5, 6) & 19.48 & 73.65 & (5, 4) & 14.07 & 43.58 & (4, 5) & 18.05 & 26.34\\
         4 & (6, 6) & 10.17 & 96.47 & (4, 5) & 8.00 & 74.16 & (5, 4) & 13.52 & 44.47 & (3, 6) & 17.37 & 27.24\\
         5 & (5, 6) & 6.47 & 96.90 & (5, 4) & 7.66 & 74.63 & (5, 4) & 12.60 & 45.25 & (4, 5) & 15.68 & 27.97\\
         6 & (5, 5) & 4.65 & 97.12 & (4, 5) & 7.00 & 75.03 & (3, 6) & 11.91& 45.96 & (4, 5) & 15.24 & 28.66\\
         7 & (4, 5) & 3.48 & 97.24 & (4, 5) & 6.67 & 75.39 & (3, 6) & 11.07 & 46.60 & (3, 6) & 14.70 & 29.32\\
         8 & (4, 5) & 3.29 & 97.35 & (5, 4) & 6.40 & 75.72 & (5, 4) & 10.56 & 47.13 & (4, 5) & 13.84 & 29.90\\
         9 & (5, 5) & 3.66 & 97.49 & (5, 4) & 6.19 & 76.03 & (5, 4) & 9.92 & 47.62 & (4, 5) & 13.76 & 30.47\\
         10 & (4, 5) & 2.92 & 97.58 & (5, 4) & 6.05 & 76.32 & (3, 6) & 9.59 & 48.08 & (2, 7) & 13.63 & 31.00\\
         \hline
         $\bm Y$ & - & - & 100 & - & - & 79.01 & - & - & 48.38 & - & - & 29.32 \\
         \hline
    \end{tabular}
    \caption{Configurations selected, the fitted $\hat\lambda_k$ and the cumulated percentage of variance explained (c.p.v.) for the first 10 iterations. Proportion of variance of the true image is shown in the bottom row}
    \label{tab:iteration}
\end{table}

 \begin{figure}[!hptb]
     \centering
     \begin{tabular}{ccccc}
          $k$ &$\sigma = 0.0$ & $\sigma = 0.1$ & $\sigma = 0.2$ & $\sigma = 0.3$\\
         1 &
         \raisebox{-.5\height}{\includegraphics[scale=0.33]{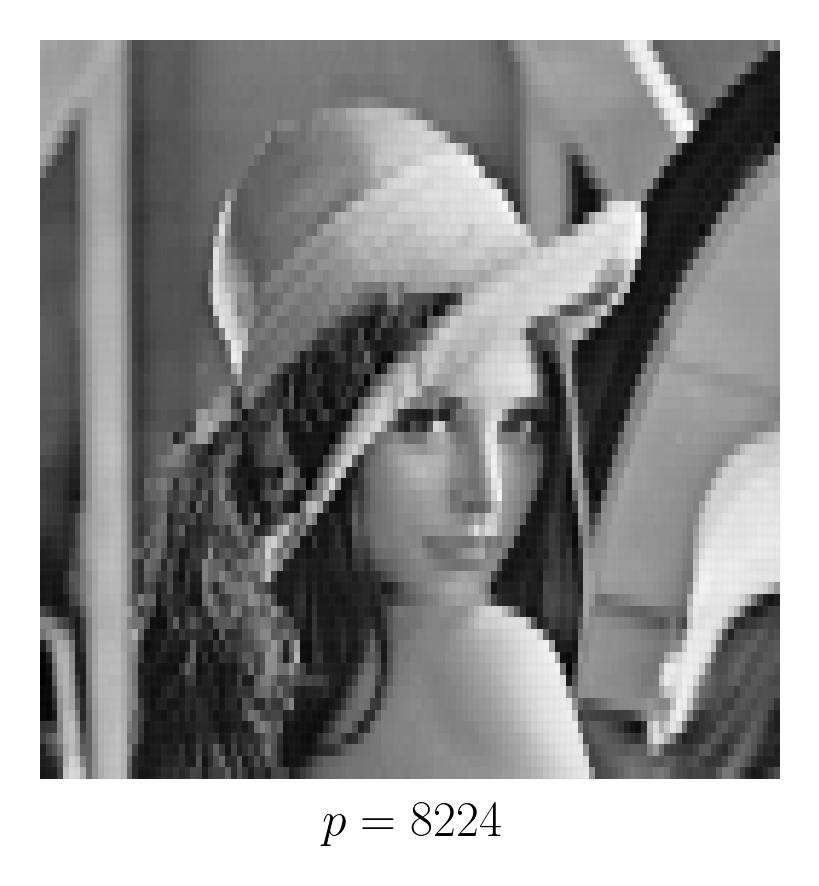}}&
         \raisebox{-.5\height}{\includegraphics[scale=0.33]{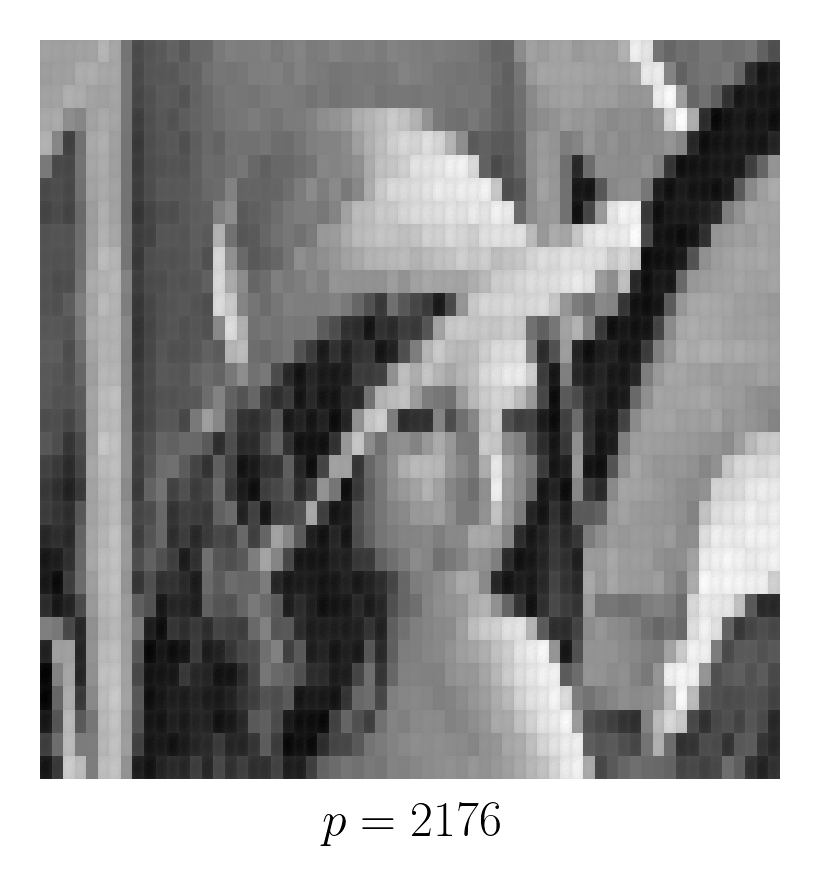}}&
         \raisebox{-.5\height}{\includegraphics[scale=0.33]{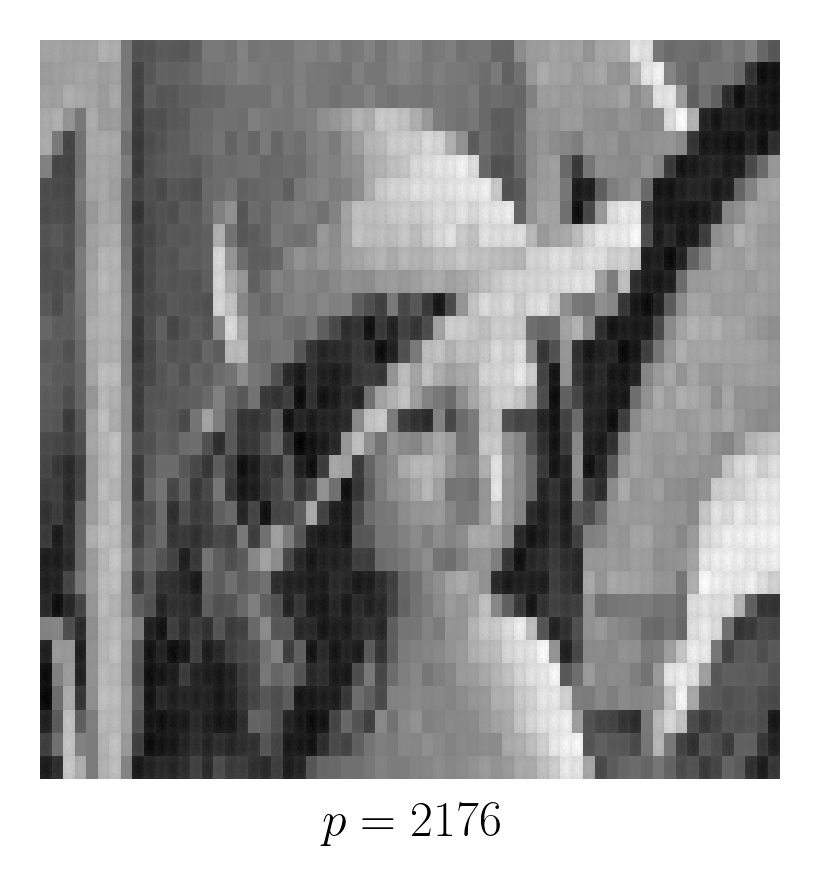}}&
         \raisebox{-.5\height}{\includegraphics[scale=0.33]{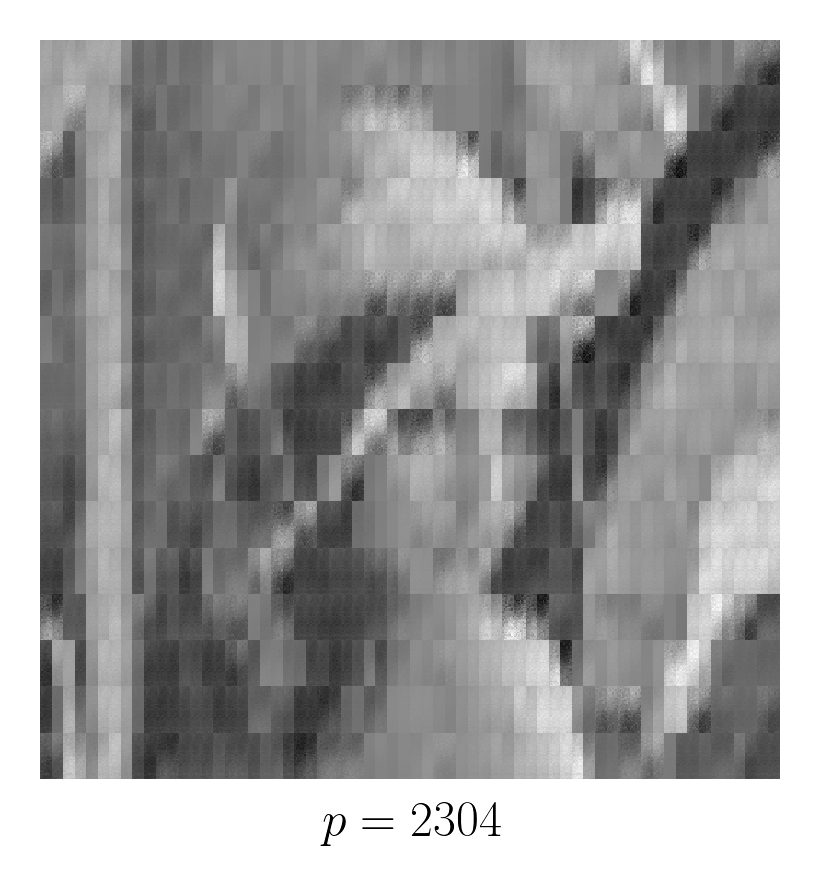}}
         \\
         2 &
         \raisebox{-.5\height}{\includegraphics[scale=0.33]{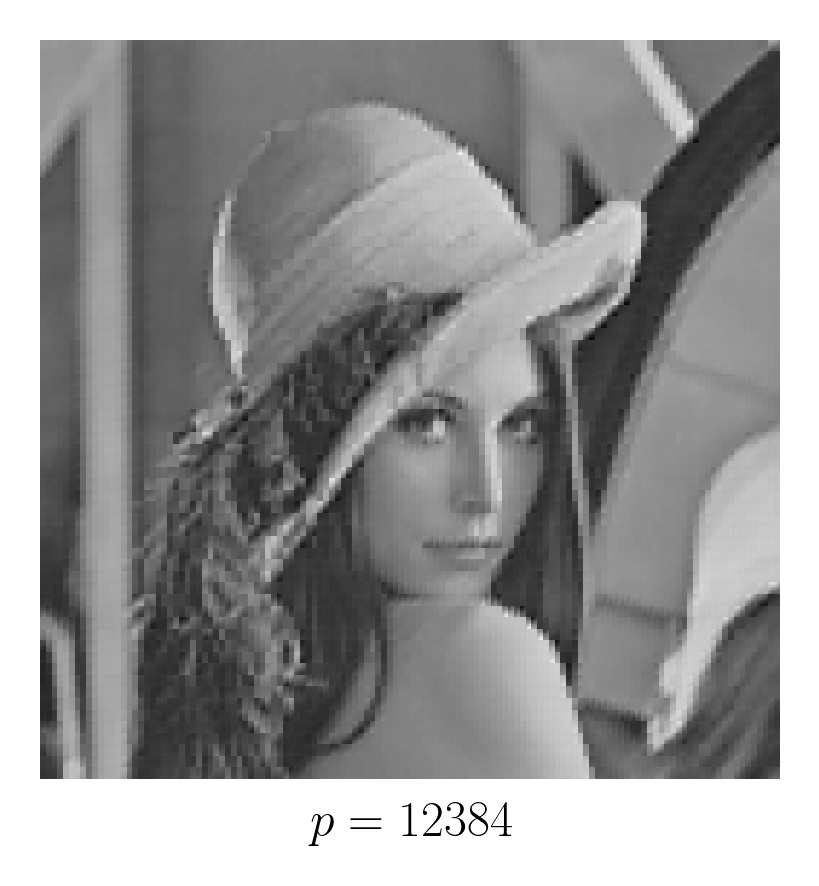}}&
         \raisebox{-.5\height}{\includegraphics[scale=0.33]{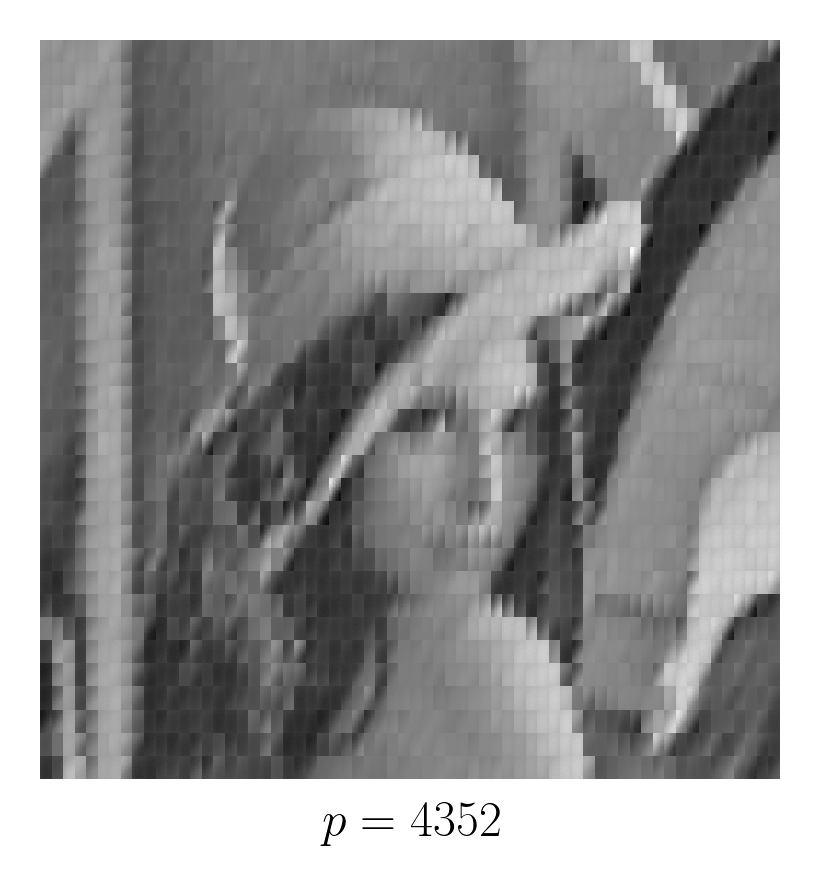}}&
         \raisebox{-.5\height}{\includegraphics[scale=0.33]{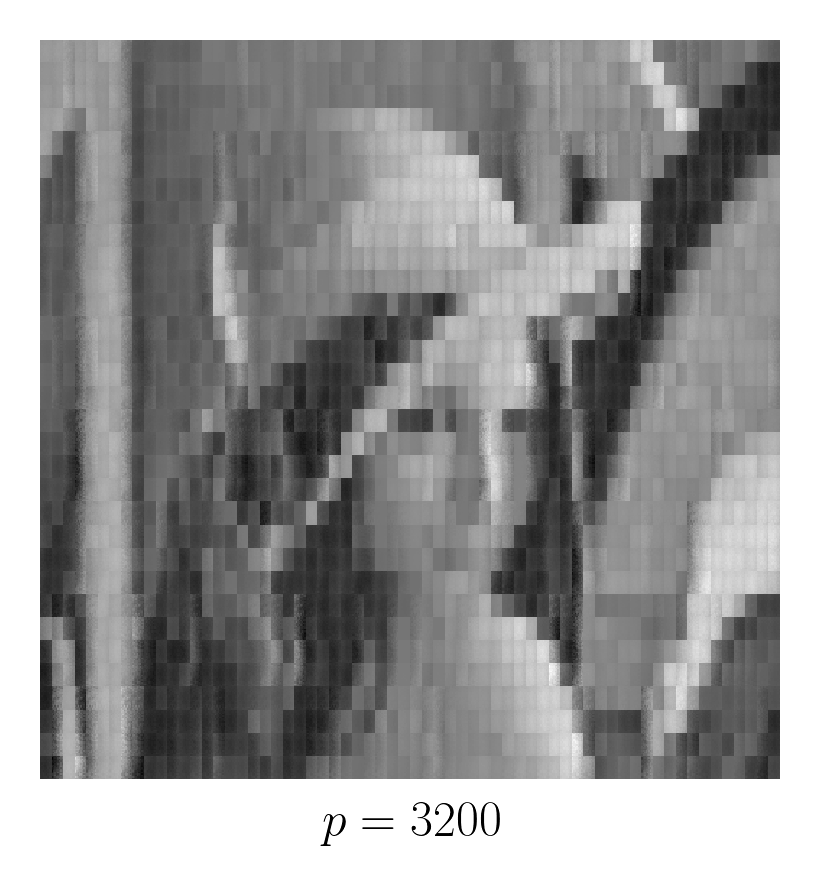}}&
         \raisebox{-.5\height}{\includegraphics[scale=0.33]{{Lenna_multi_term_noise_0.3_kpd_2_p_2304}.png}}
         \\
         3 &
         \raisebox{-.5\height}{\includegraphics[scale=0.33]{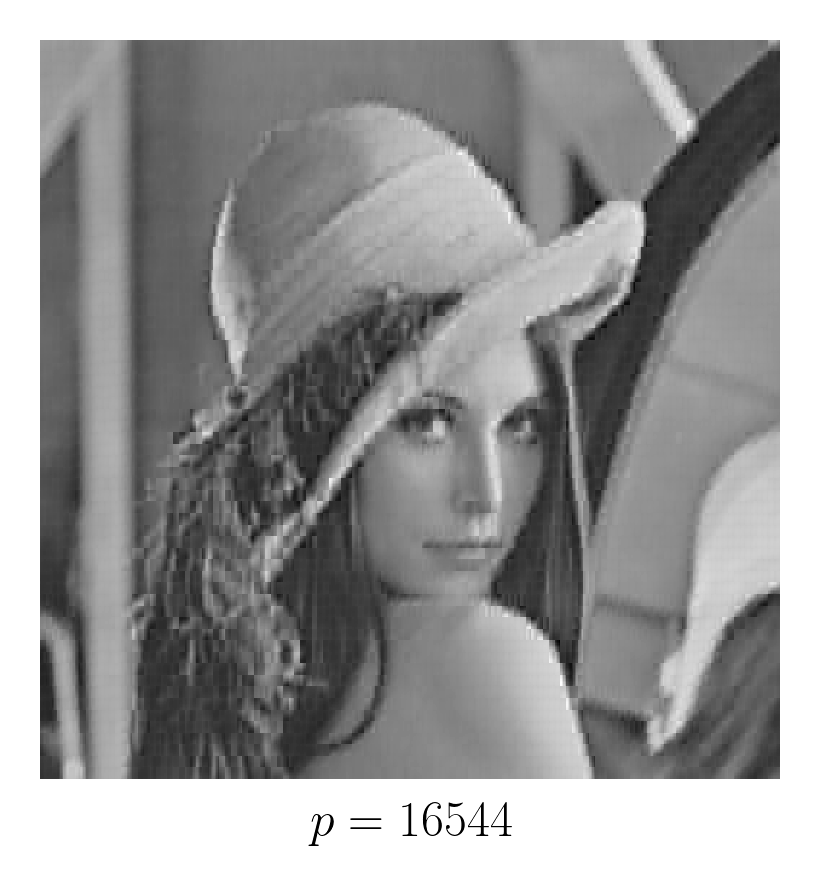}}&
         \raisebox{-.5\height}{\includegraphics[scale=0.33]{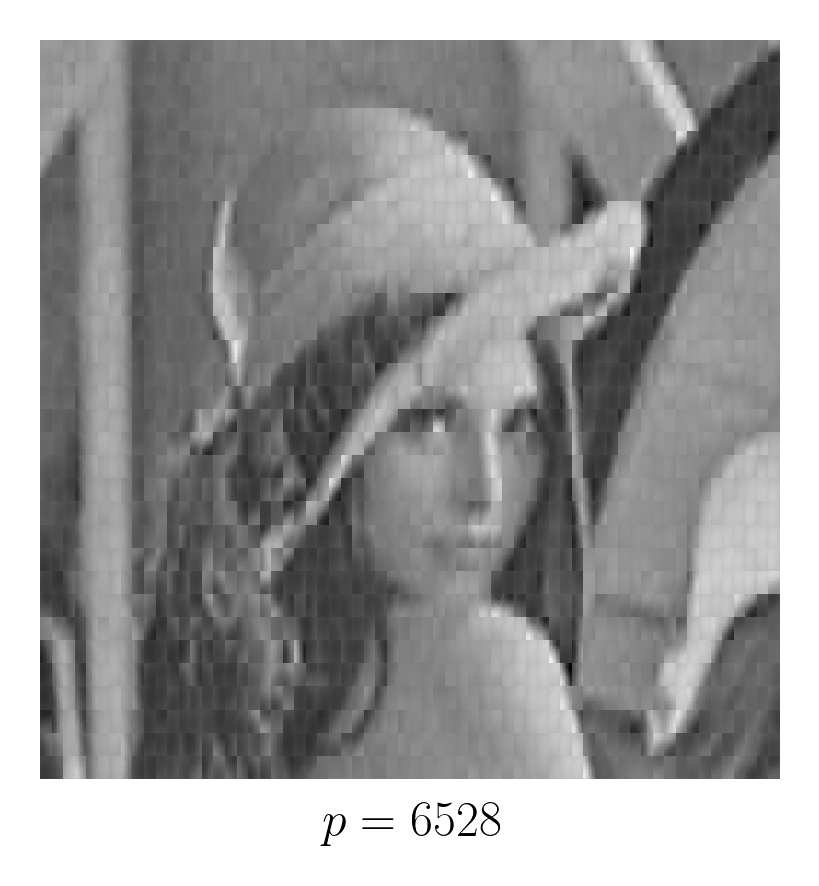}}&
         \raisebox{-.5\height}{\includegraphics[scale=0.33]{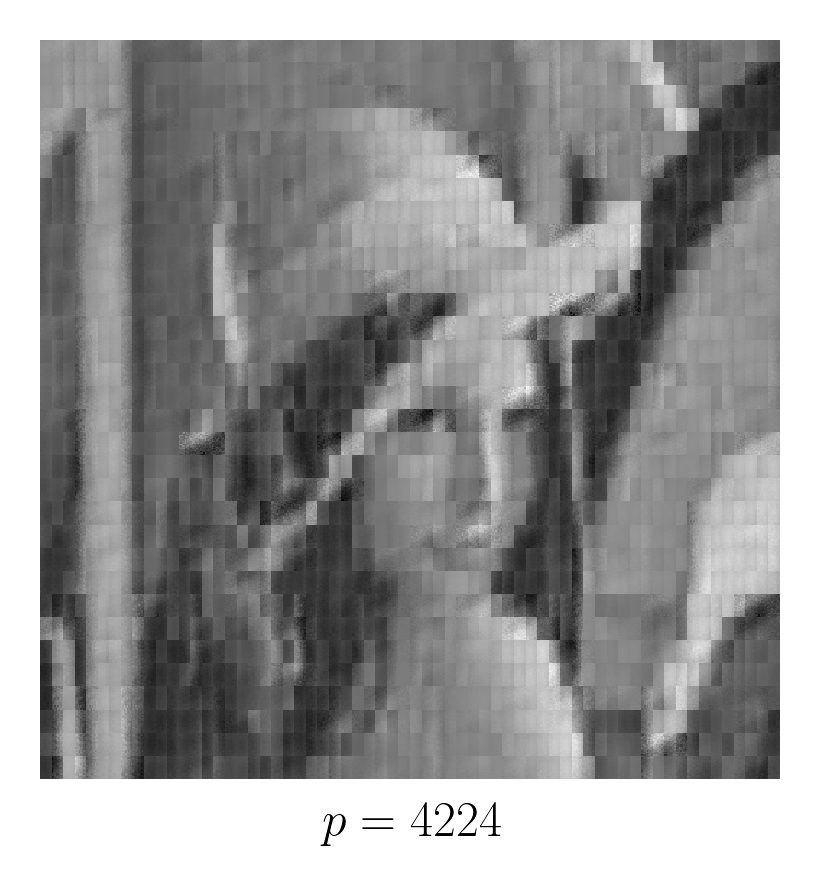}}&
         \raisebox{-.5\height}{\includegraphics[scale=0.33]{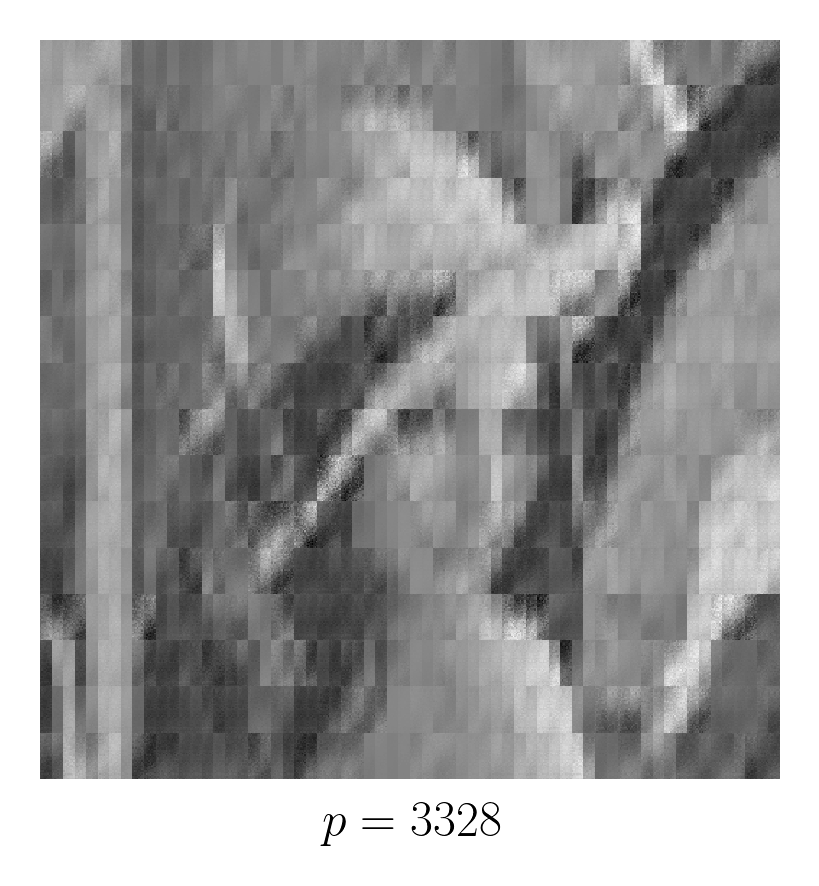}}
         \\
         4 &
         \raisebox{-.5\height}{\includegraphics[scale=0.33]{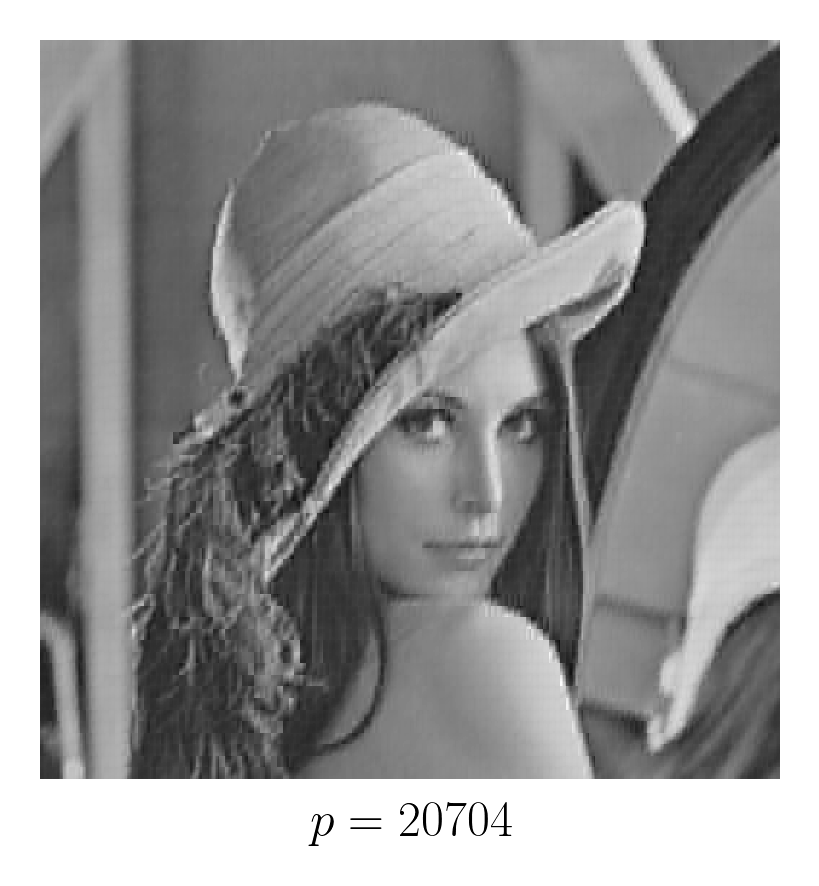}}&
         \raisebox{-.5\height}{\includegraphics[scale=0.33]{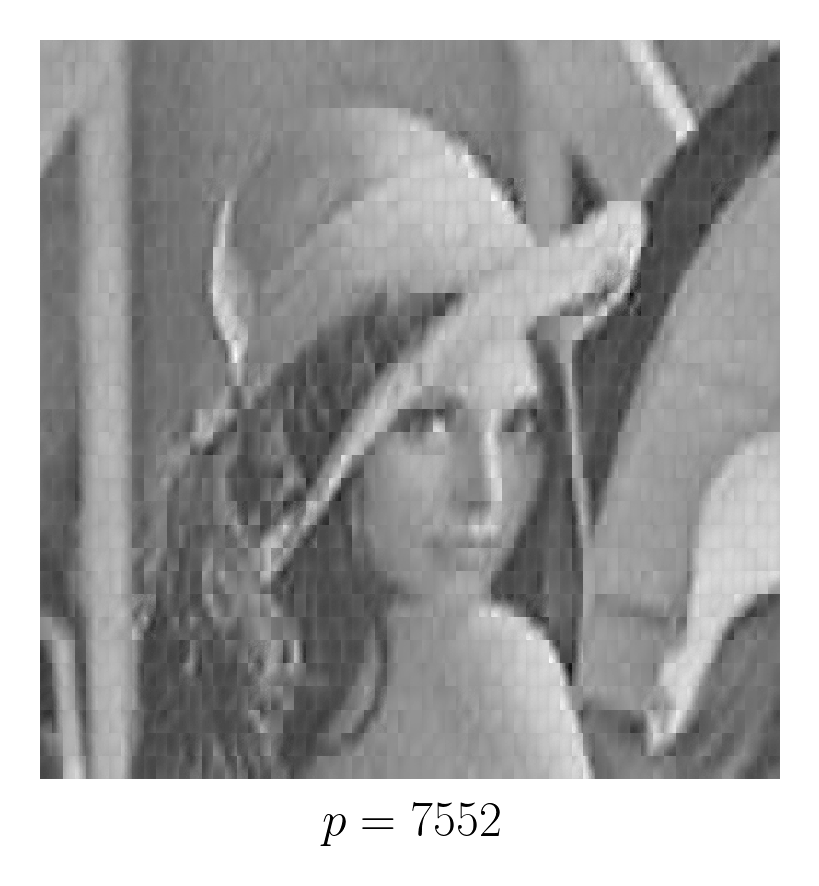}}&
         \raisebox{-.5\height}{\includegraphics[scale=0.33]{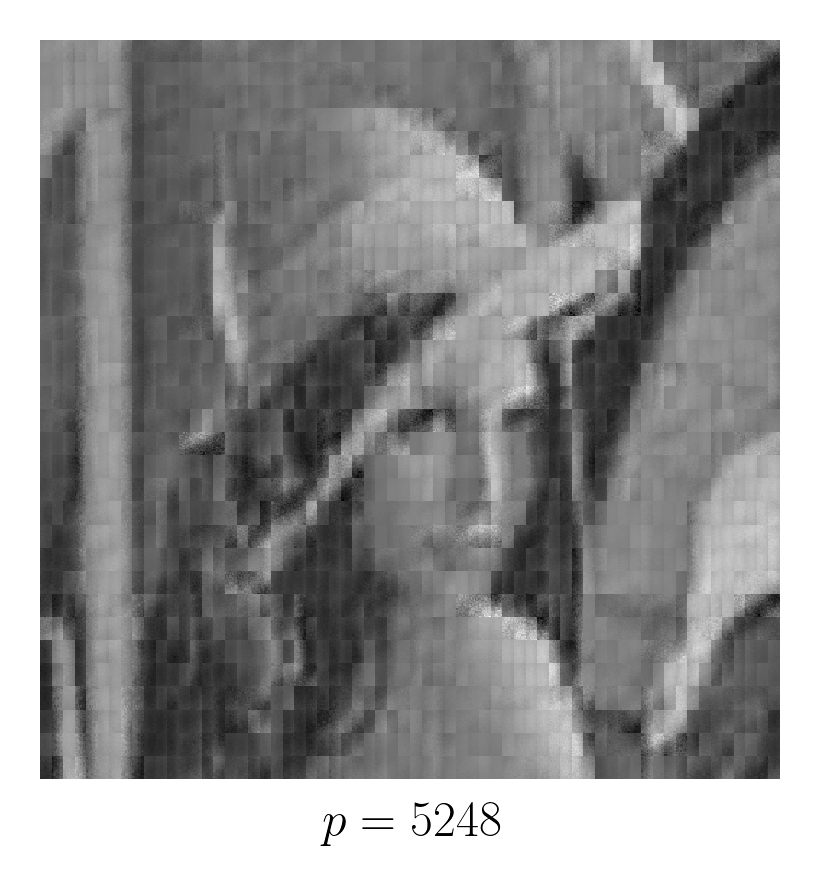}}&
         \raisebox{-.5\height}{\includegraphics[scale=0.33]{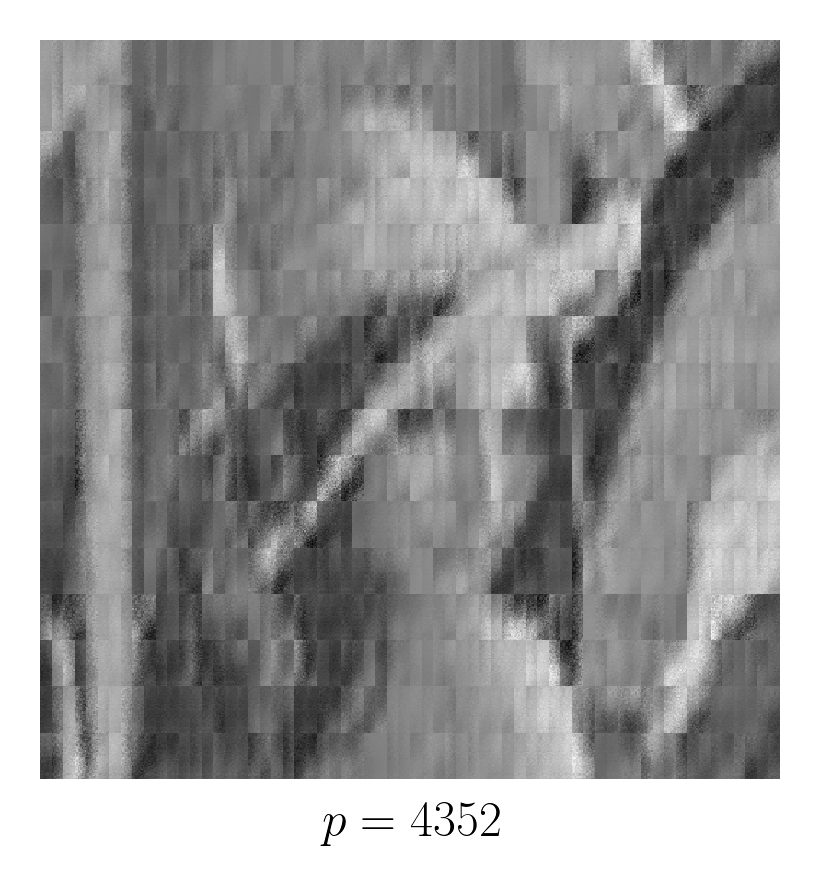}}
         \\
         5 &
         \raisebox{-.5\height}{\includegraphics[scale=0.33]{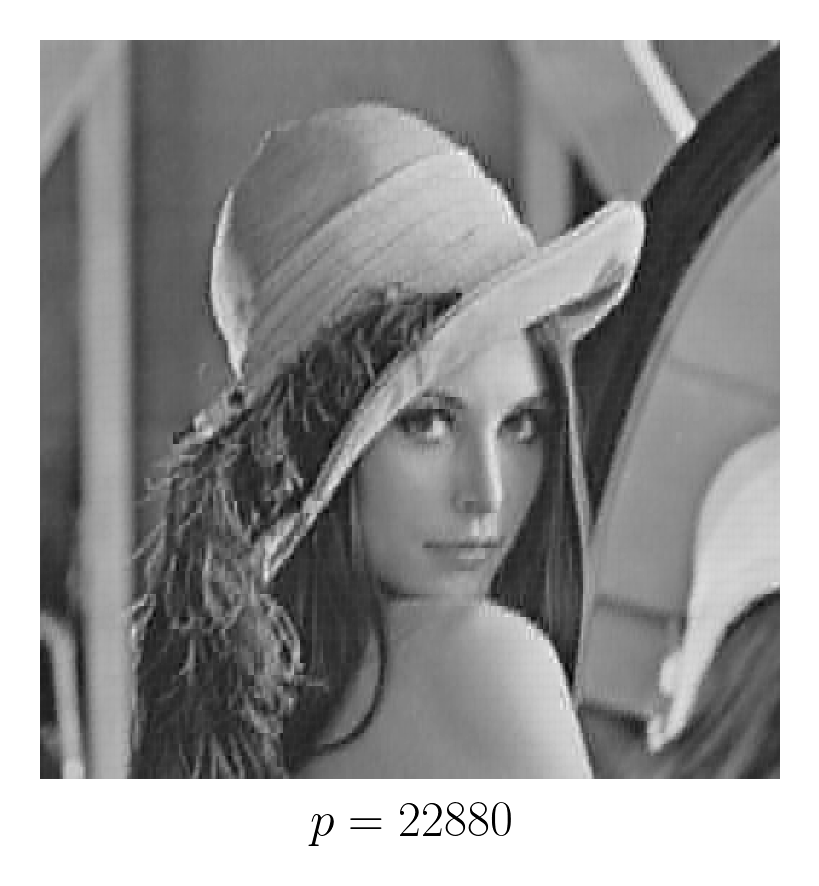}}&
         \raisebox{-.5\height}{\includegraphics[scale=0.33]{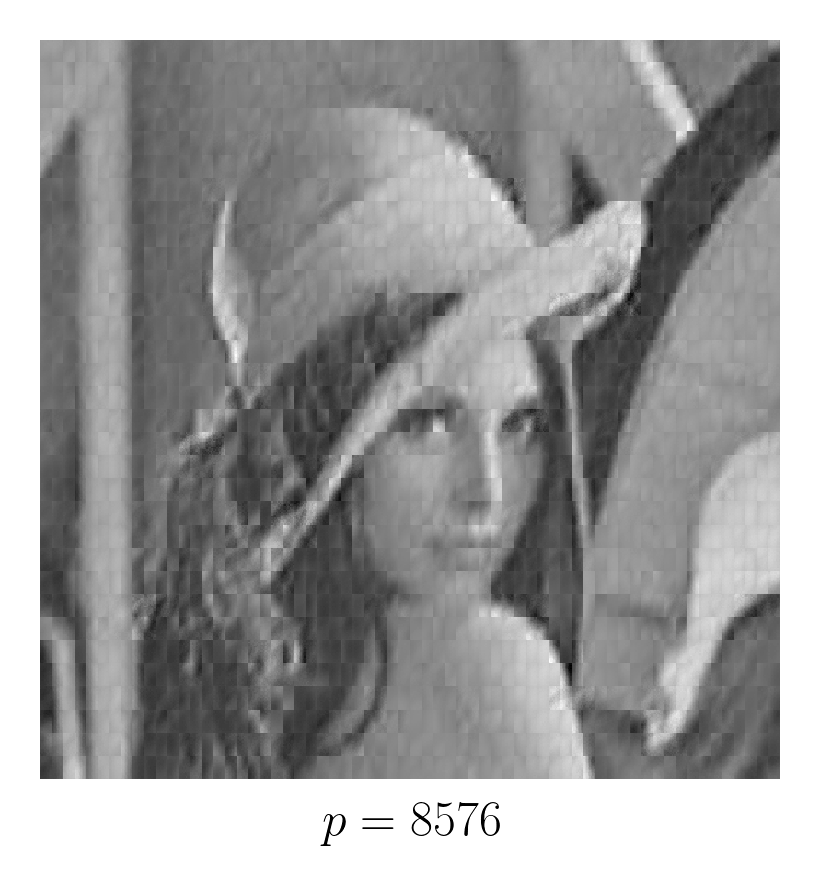}}&
         \raisebox{-.5\height}{\includegraphics[scale=0.33]{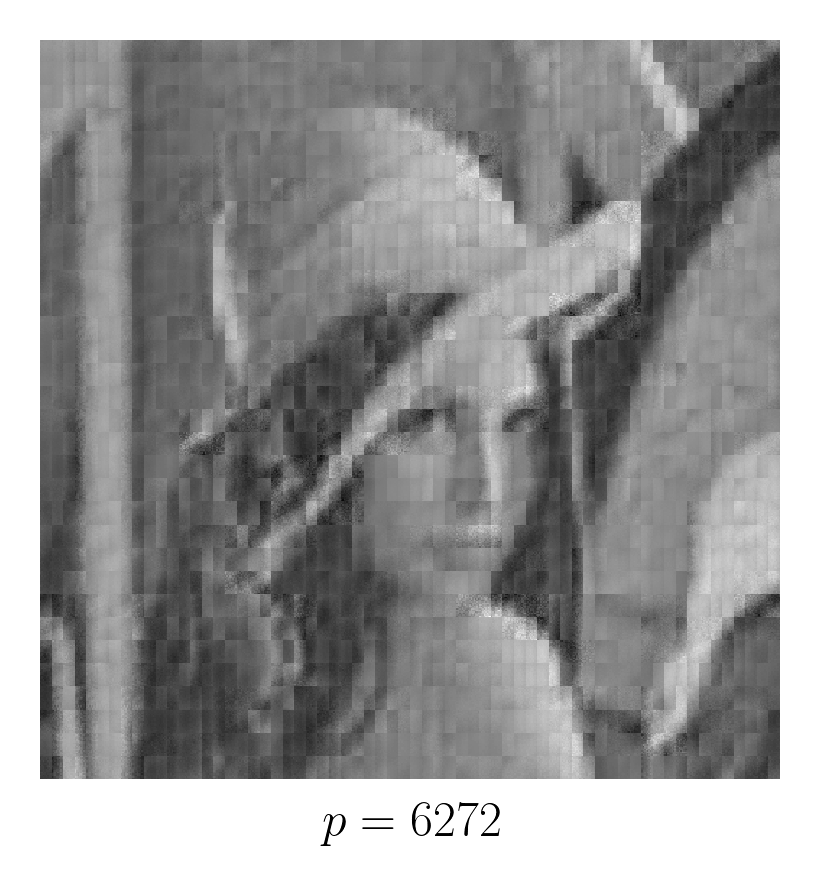}}&
         \raisebox{-.5\height}{\includegraphics[scale=0.33]{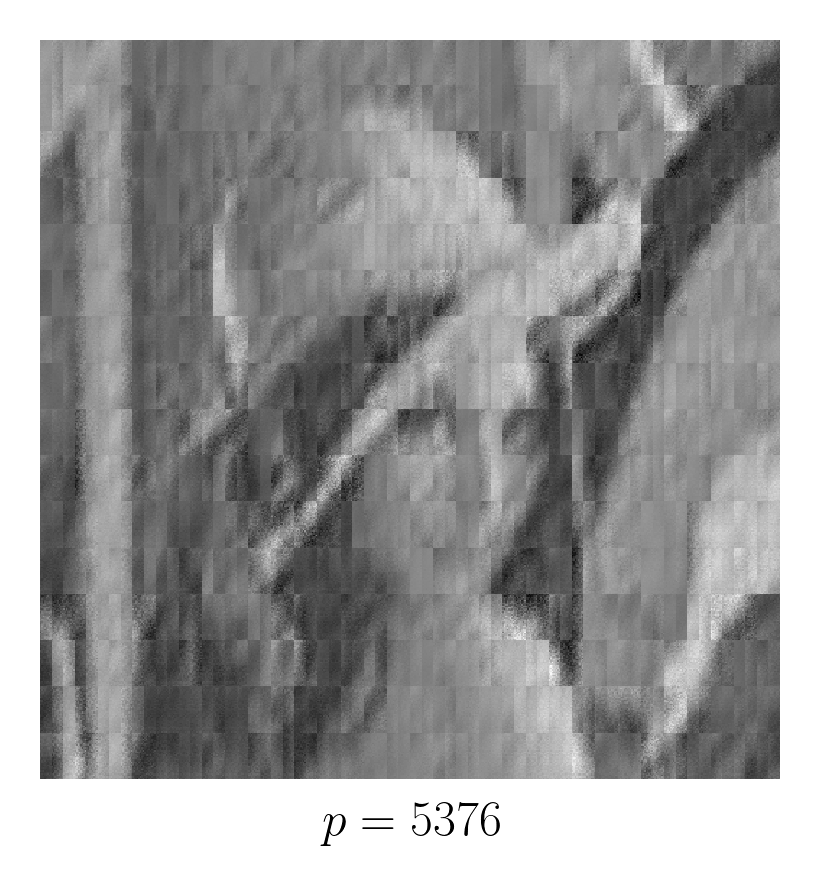}}
         \\
         6 &
         \raisebox{-.5\height}{\includegraphics[scale=0.33]{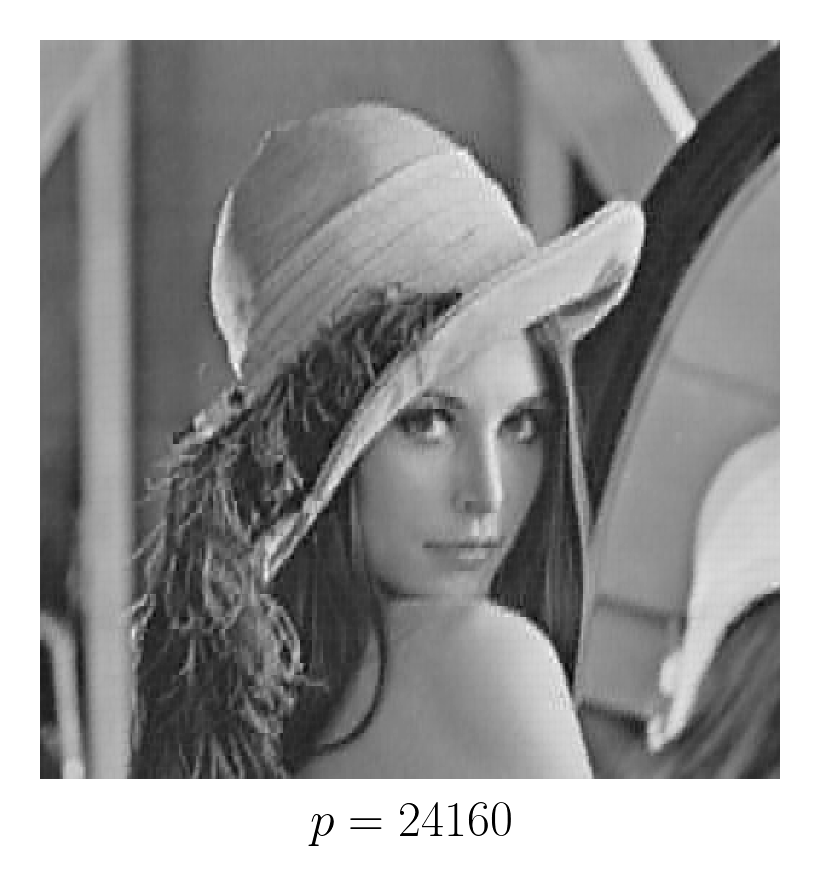}}&
         \raisebox{-.5\height}{\includegraphics[scale=0.33]{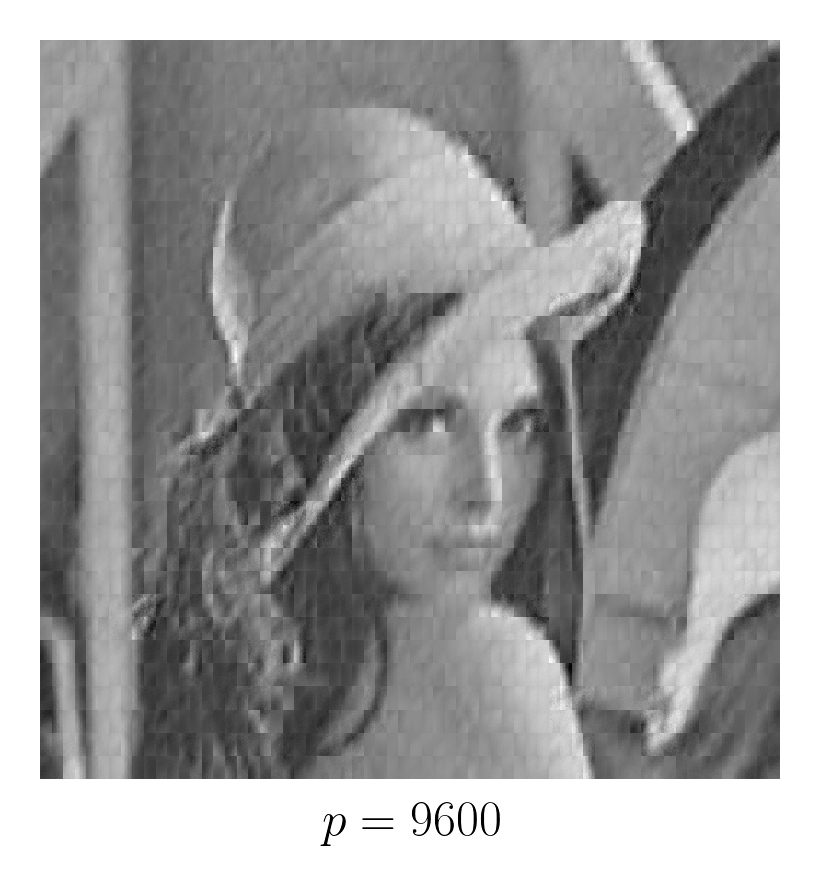}}&
         \raisebox{-.5\height}{\includegraphics[scale=0.33]{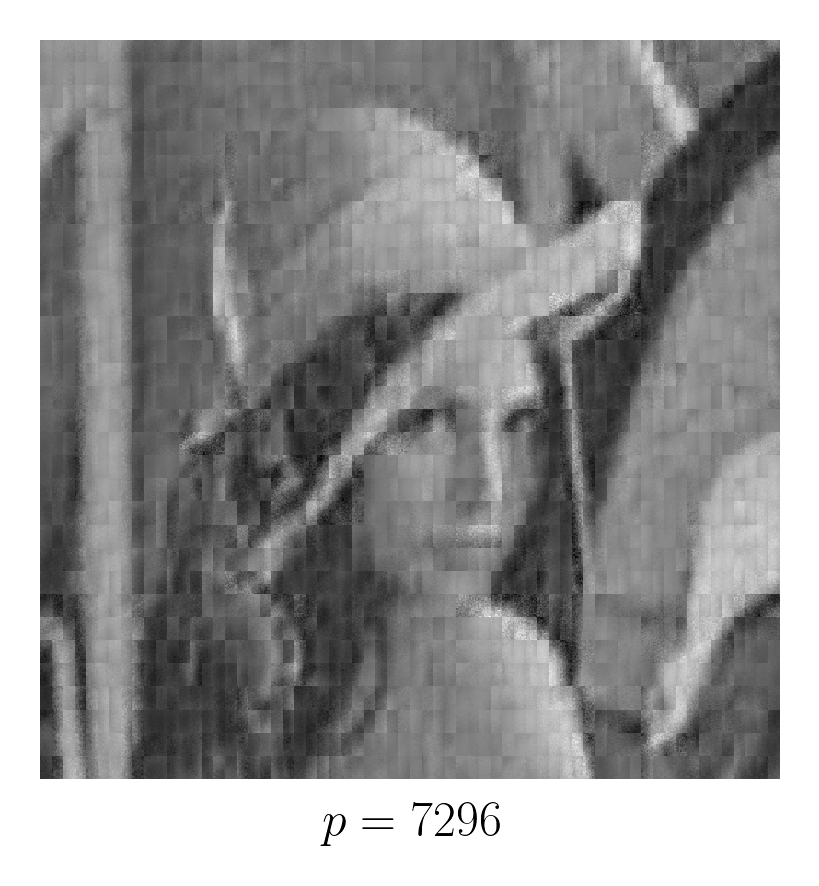}}&
         \raisebox{-.5\height}{\includegraphics[scale=0.33]{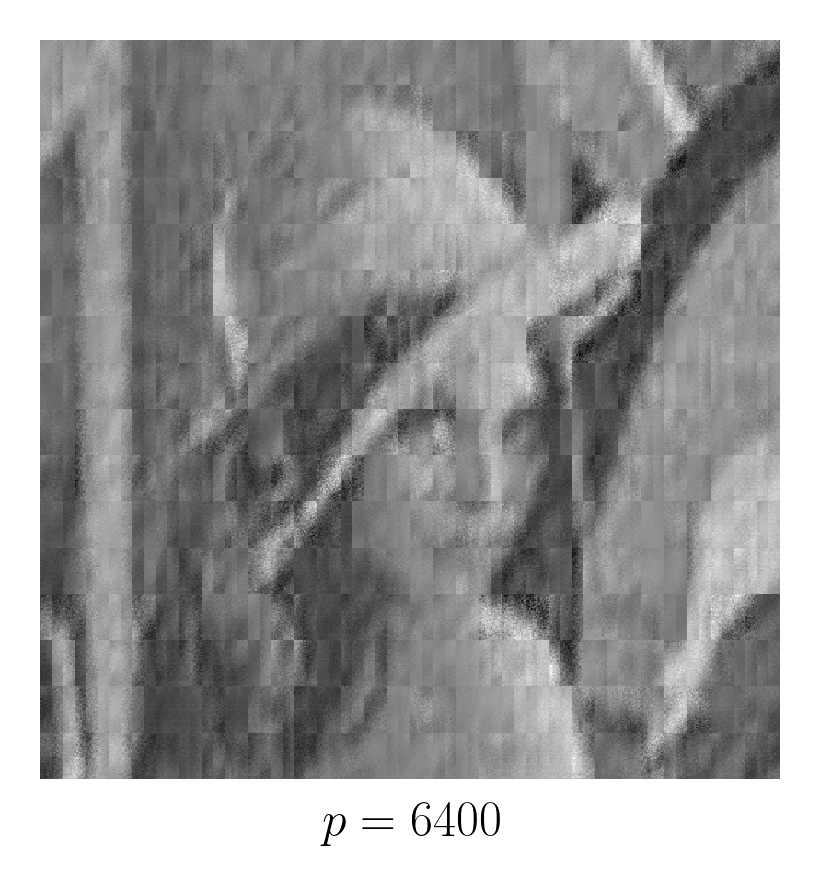}}
         \\
     \end{tabular}
     \caption{Fitted Images in the first 6 iterations.}
     \label{fig:fitted-image}
 \end{figure}

 The fitted images of using one to six Kronecker product terms at different noise levels $\sigma$ are plotted in Figure \ref{fig:fitted-image} with the number of parameters shown under each image. It is seen from the figures that, as expected, the smaller noise levels result in better recovery of the image with less numbers of Kronecker product terms. Note that the images plotted in Figure \ref{fig:fitted-image} are ordered according to the number of
 Kronecker product terms used instead of the number of parameters used. The configuration $(6, 7)$ selected in the first iteration in the $\sigma = 0$ case
 uses roughly four times the number of parameters of the configuration $(5, 6)$ selected in the
 $\sigma = 0.1$ case.

In additional to the iterative greedy algorithm, we use a stopping criterion based on random matrix theory to determine number of Kronecker products. Specifically, at iteration $k$, an estimate of $\sigma$ is 
$$\hat\sigma = 2^{-(M+N)/2}\|\hat{\bm E}^{(k)} - \hat {\bm S}_k\|_F = 2^{-(M+N)/2}\|\hat{\bm E}^{(k+1)}\|_F.$$
Under the i.i.d. Gaussian assumption on $\hat{\bm E}^{(k+1)}$, we have
$$P\left[\|\mathcal R_{\hat m_k, \hat n_k}[\hat{\bm E}_{k+1}]\|_S\geqslant \hat\sigma \left(2^{(\hat m_k+\hat n_k)/2}+2^{(M+N-\hat m_k-\hat n_k)/2}+t\right)\right]\leqslant e^{-t^2/2},$$
according to the non-asymptotic analysis on the random matrices \citep{Vershynin2010Introduction}.
Here we set $t = \sqrt{2\log 100}\approx 3.03$ such that the probability is bounded by 0.01. We terminate the algorithm at step $k$ if 
\begin{equation}
\hat\lambda_k \leqslant \hat\sigma \left(2^{(\hat m_k+\hat n_k)/2}+2^{(M+N-\hat m_k-\hat n_k)/2}+\sqrt{2\log 100}\right),\label{eq:stopping}
\end{equation}
and use the first $\hat K=k-1$ terms as the optimal approximation.
Specifically, when $\sigma = 0$ or $0.1$, the stopping criterion is never met in the first $20$ iterations and we use $\hat K = 20$. When $\sigma = 0.2$, a 9-term model is selected, and when $\sigma =0.3$, the stopping criterion results in a 7-term model.

\begin{figure}[!htb]
    \centering
    \includegraphics[scale=0.5]{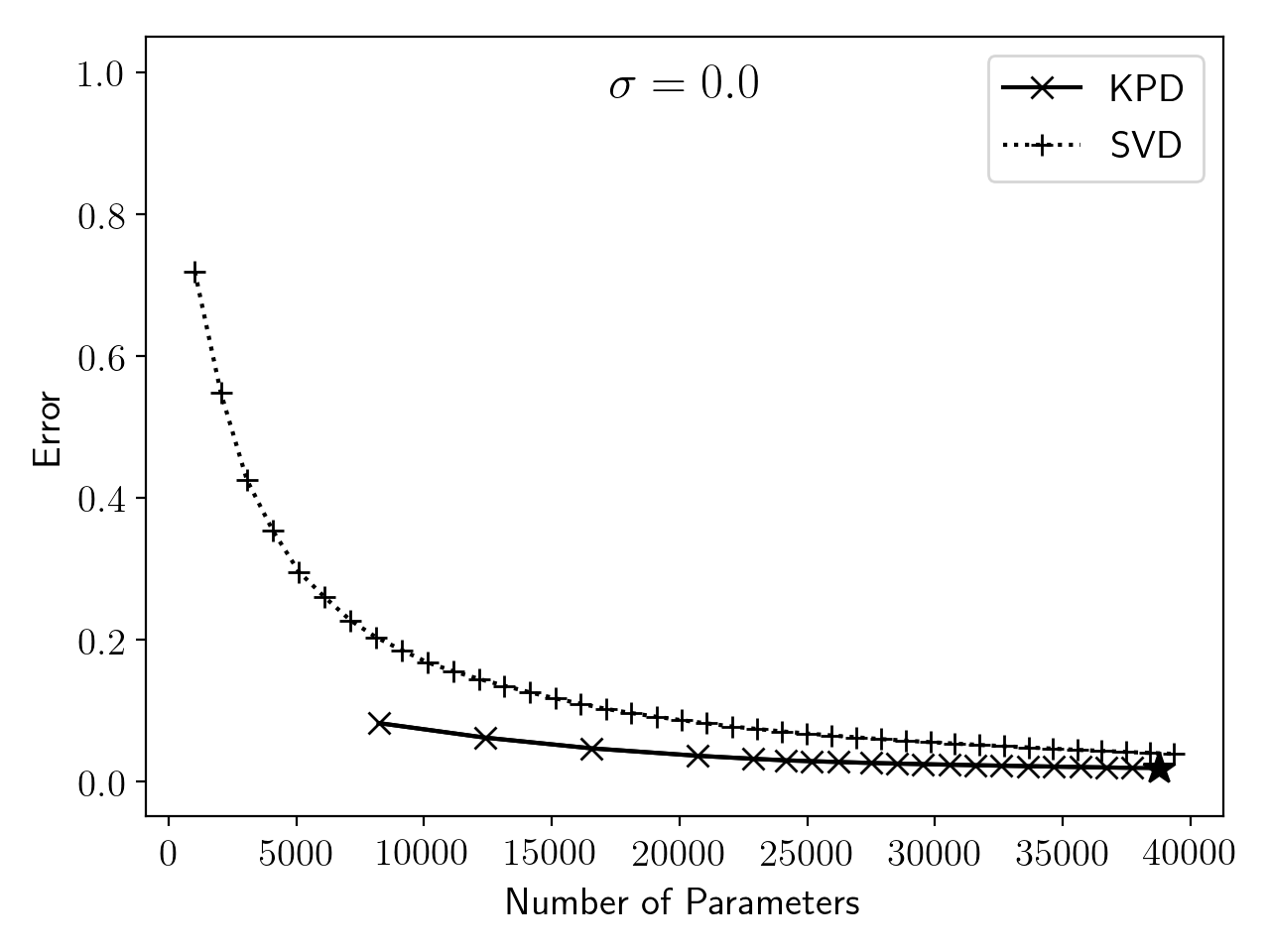}
    \includegraphics[scale=0.5]{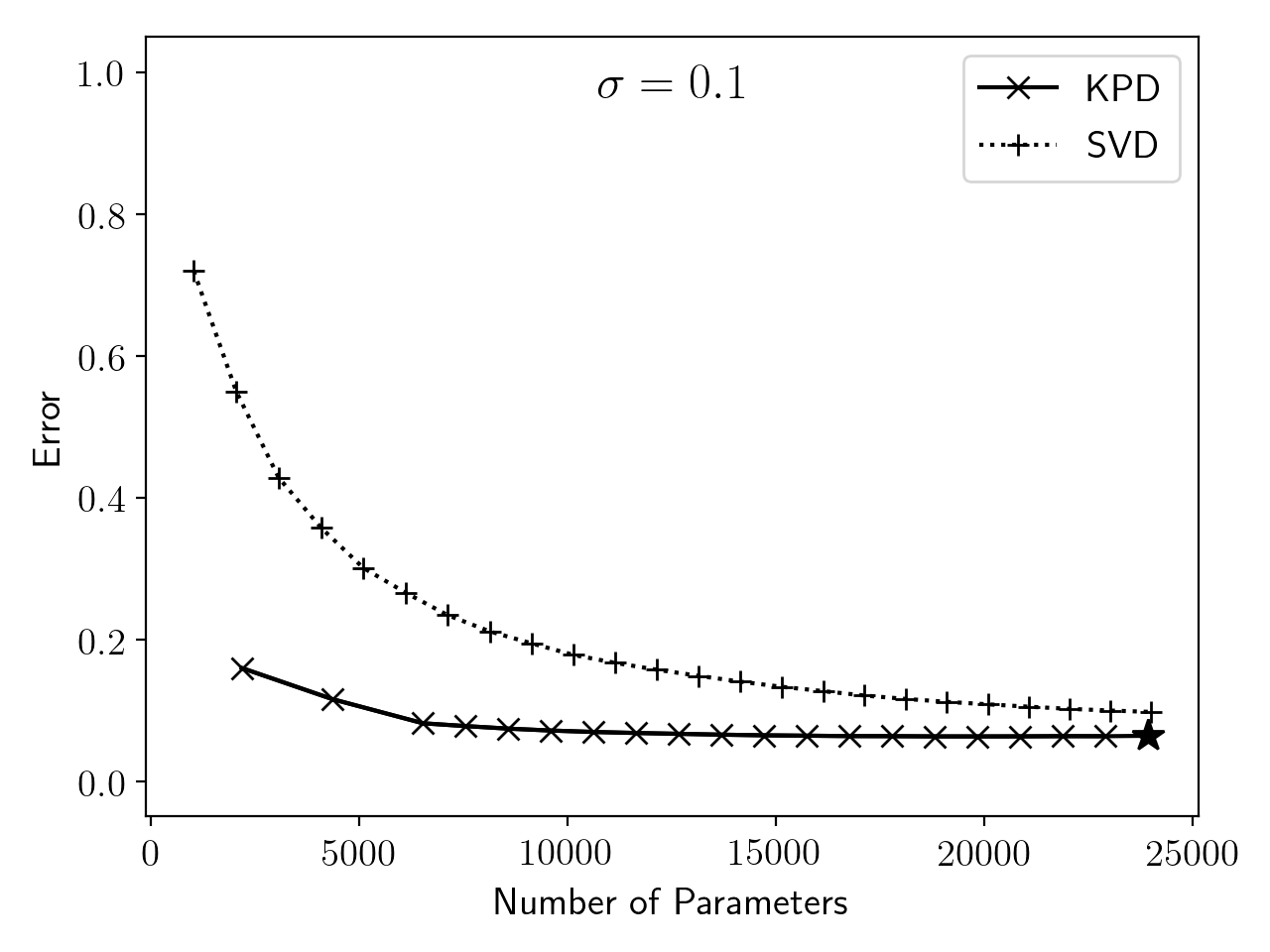}\\
    \includegraphics[scale=0.5]{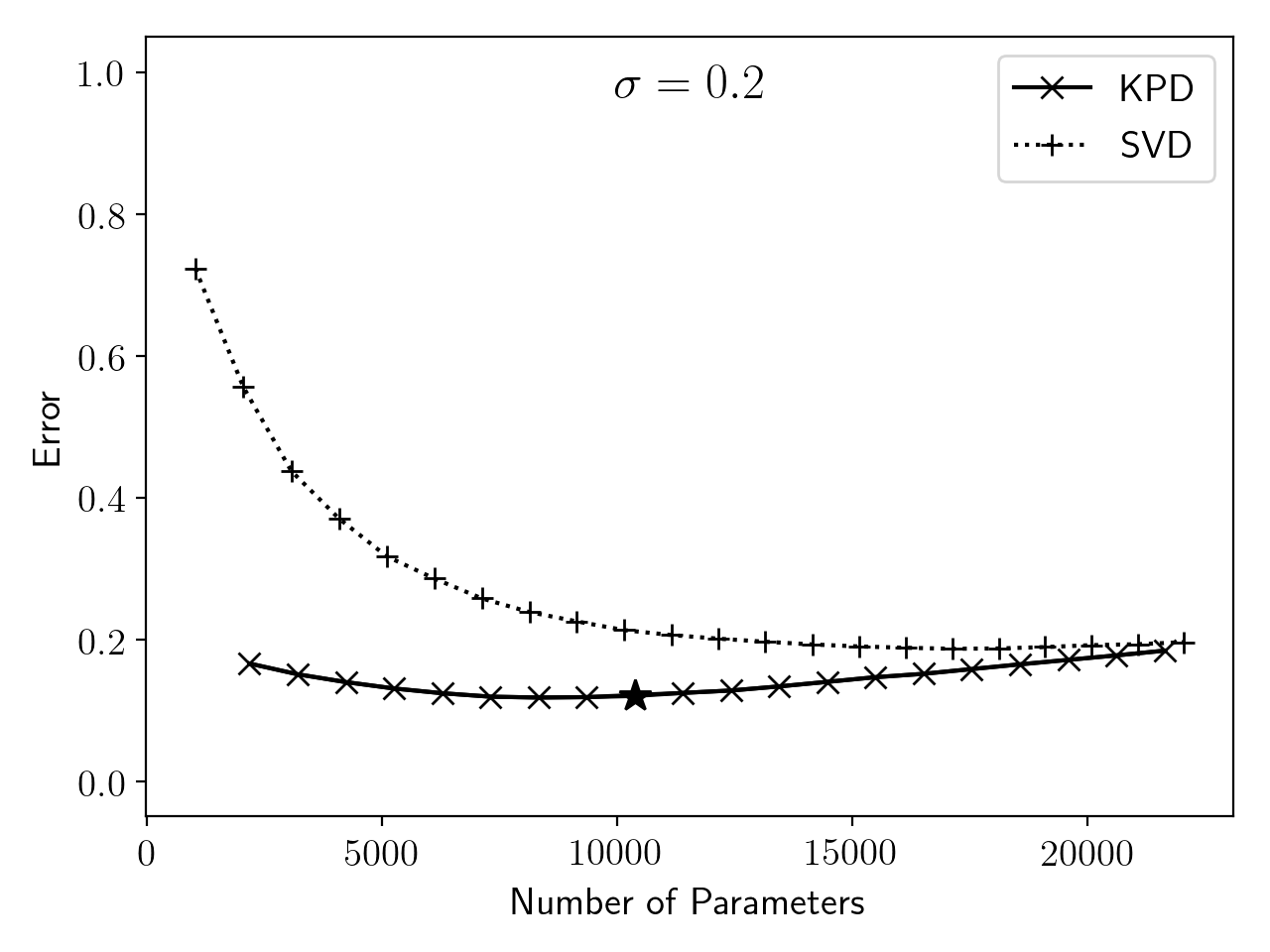}
    \includegraphics[scale=0.5]{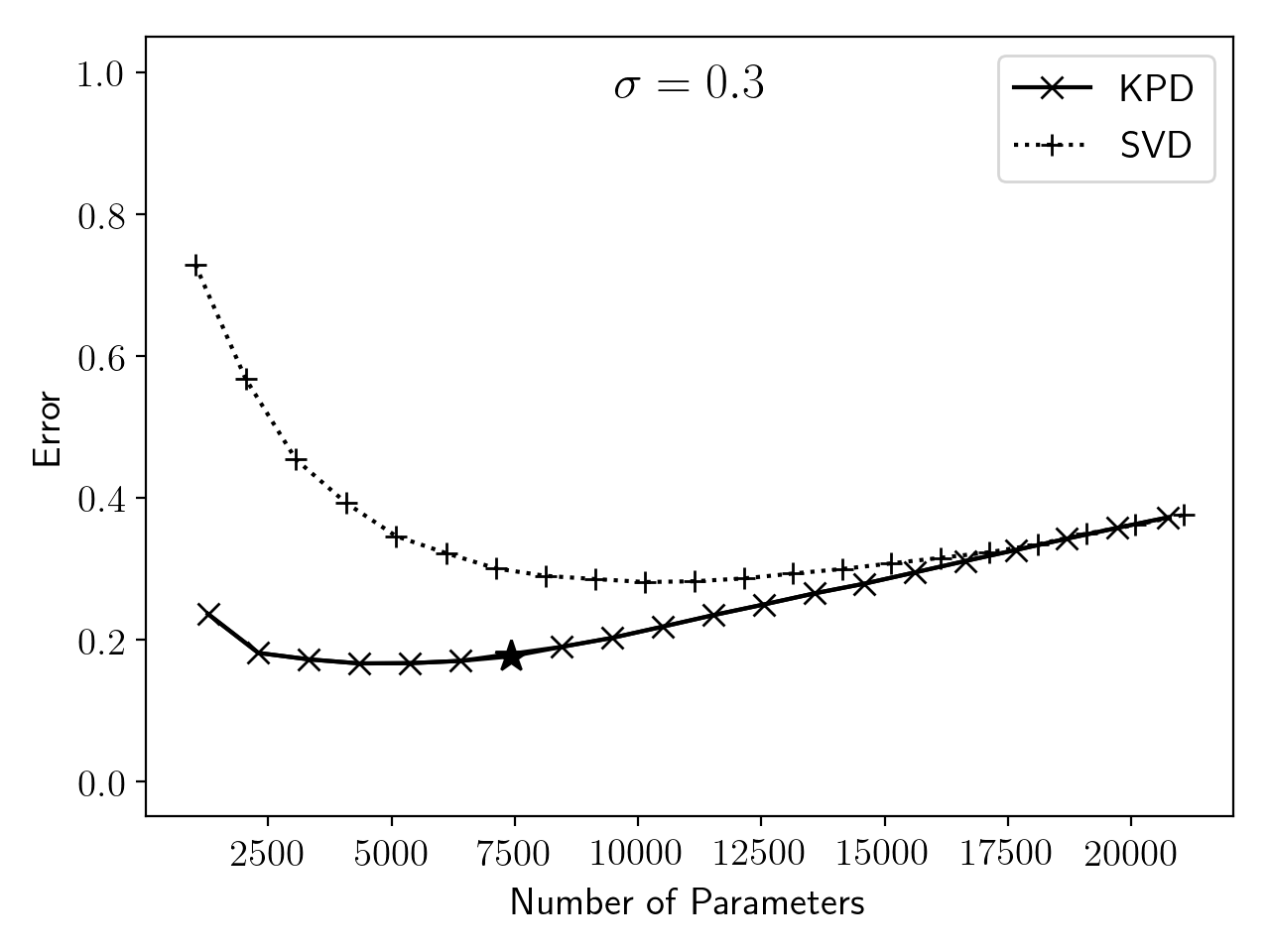}
    \caption{Error of fitted matrix against the number of parameters used.}
    \label{fig:error-curve}
\end{figure}

 Singular value decomposition (SVD) is another widely used approach in image denoising and compression. It assumes
 \[
 \bm Y=\sum_{i=1}^K\lambda_i \bm u_i\bm v_i^T.
 \]
 The complexity level in such a approach is $K$, the number of rank one matrices used in the approximation. Note that SVD is a special case of $h$KoPA model in which each Kronecker product is a vector outer product, corresponding to configuration $(M,1)$ in KoPA model.
 Hence no configuration determination
 is needed. The number of parameters used in such an approximation is
 $K(2^M+2^N-1)$ where $K$ is the number of rank-one matrices used.
 To compare the performance of $h$KoPA with the SVD approach,
 we calculate the relative squared error of the fitted matrix $\hat{\bm Y}$ by
 $$
 RSE=\dfrac{\|\bm Y - \hat{\bm Y}\|_F^2}{\|\bm Y\|_F^2},
 $$
 where $\bm Y$ is the original image without noise and $\hat{\bm Y}$ is the
 fitted matrix of the noisy version $\bm Y_\sigma$.
 For different noise levels $\sigma$, we plot RSE against the
 number of parameters used in the approximation, shown in Figure \ref{fig:error-curve} for both SVD approximation and $h$KoPA. Models with different number of terms are marked in Figure \ref{fig:error-curve} and the $h$KoPA with the proposed random matrix stopping criterion is highlighted with a ``$\bm\star$" marker.
 Comparing the error curve of SVD with the one of $h$KoPA, Figure \ref{fig:error-curve} reveals that for any level of model complexity (or the number of parameters), $h$KoPA is
 more accurate than the standard low rank SVD approximation. When noises are added,
 overfitting is observed for both $h$KoPA and SVD approximation as
 the error (comparing to the true image) increases when too many terms are
 used, seen from the $U$-shape of the curves. The stopping criterion in (\ref{eq:stopping}) prevents the model from significantly overfitting. The realized relative error of the $h$KoPA with number of terms selected by (\ref{eq:stopping}) is close to the minimum attainable error, though the estimated number of terms is not the optimal one. 

\newpage
\section{Conclusion and Discussion}\label{sec:conclusion}
In this paper, we extend the single-term KoPA model proposed in \cite{cai2019kronecker} to a more flexible setting, which allows multiple terms with different configurations and allows the configurations to be unknown.
Identifiability conditions are studied to ensure unique representation of the
model. And we propose two iterative estimation algorithms.

With a given set of configurations,
we propose a least squares backfitting algorithm that fits each Kronecker product component in an alternating way.
The simulation study shows the performance of the algorithm and the
impact of the linear dependency between the component matrices.

When the configurations are unknown, the extra flexibility of $h$KoPA
allows for more parsimonious representation of the underlying matrix, though
it comes with the challenge of configuration determination. An iterative greedy
algorithm is proposed to jointly determine the configurations and estimate
each Kronecker product component. The algorithm iteratively adds one Kronecker
product term to
the model by finding the best one term KoPA to the residual matrix obtained
from the previous iteration, using the procedure proposed in
\cite{cai2019kronecker}.
By analyzing a real image, we demonstrated that the proposed algorithm
is able to obtain reasonable $h$KoPA and the results are significantly superior
than the standard low rank matrix approximation using SVD.

The matrix dimensions discussed in this article are powers of 2. This
is certainly not necessary. If the observed matrix is dimension $P\times Q$,
and if $P$ has factors $1=p_0<p_1<\ldots<p_M=P$ and
$Q$ has factors $1=q_0<q_1<\ldots<q_N=Q$, then the possible configuration set
is all combinations of $(p_m,q_n)$ for $0\leq m\leq M$ and $0\leq n\leq N$,
excluding the trivial $(1,1)$ and $(P,Q)$ cases. Note that $(P,1)$
configuration corresponds to the
traditional low rank matrix SVD approximation setting.
All algorithms presented in this paper apply for the general $P\times Q$ cases.
Of course the more factors
$P$ and $Q$ have, the larger the possible configuration set is hence we have
more
flexibility to find a better approximation. One can augment the
observed matrix with additional rows and columns to increase the
size of allowable configuration set. In image processing,
a common tool used for augmentation is
super-sampling by interpolating more pixels. In matrix denoising and
compression, one can add zeros or duplicate certain rows and columns
of the original matrix.

As discussed in Section 3, the greedy algorithm for configuration determination
is similar to boosting algorithm. It can be improved at the expense of higher
computational cost. In addition to combine it with the backfitting estimation
procedure under a fixed configuration obtained by the greedy algorithm, the
configurations can also be improved and adjusted
under the backfitting framework. Their theoretical properties need to be
investigated. A more sophisticated stopping criterion also needs to be
investigated. Rrevious researches on principle component analysis rank determination such as those \cite{minka2001automatic, lam2012factor, bai2018consistency} may be extended to our case.


\newpage
\bibliographystyle{apa}
\bibliography{references}
\end{document}